\documentclass[11pt]{article}

\usepackage[utf8]{inputenc}
\usepackage{amsmath, amssymb}
\usepackage{graphicx}
\usepackage{url}
\usepackage{authblk}
\usepackage{natbib}
\usepackage{geometry}
\usepackage{hyperref}
\usepackage{algorithm}
\usepackage{algpseudocode}
\usepackage{booktabs}
\usepackage{multirow}

\hypersetup{
    colorlinks=true,
    linkcolor=blue,
    citecolor=blue,
    urlcolor=blue
}

\title{The Impact of Climatic Factors on Respiratory Pharmaceutical Demand: A Comparison of Forecasting Models for Greece}

\author{Viviana Schisa\thanks{Corresponding author: viviana.schisa2@unibo.it}}
\author{Matteo Farn\`e}

\affil{Department of Statistical Sciences, Alma Mater Studiorum - University of Bologna, Bologna, Italy}

\date{}

\begin{document}

\maketitle

\begin{abstract}
Climate change is increasingly recognized as a driver of health-related outcomes, yet its impact on pharmaceutical demand remains largely understudied. As environmental conditions evolve and extreme weather events intensify, anticipating their influence on medical needs is essential for designing resilient healthcare systems.

This study examines the relationship between climate variability and the weekly demand for respiratory prescription pharmaceuticals in Greece, based on a dataset spanning seven and a half years (390 weeks). Granger causality spectra are employed to explore potential causal relationships. Following variable selection, four forecasting models are implemented: Prophet, a Vector Autoregressive model with exogenous variables (VARX), Random Forest with Moving Block Bootstrap (MBB-RF), and Long Short-Term Memory (LSTM) networks.

The MBB-RF model achieves the best performance in relative error metrics while providing robust insights through variable importance rankings. The LSTM model outperforms most metrics, highlighting its ability to capture nonlinear dependencies. The VARX model, which includes Prophet-based exogenous inputs, balances interpretability and accuracy, although it is slightly less competitive in overall predictive performance.

These findings underscore the added value of climate-sensitive variables in modeling pharmaceutical demand and provide a data-driven foundation for adaptive strategies in healthcare planning under changing environmental conditions.
\end{abstract}

\section{Introduction}

Climate change represents one of the foremost challenges of recent decades, with profound impacts on environmental systems, public health, and the global economy.
This phenomenon is projected to exacerbate global inequality and public health risks, pushing more than 100 million into poverty and limiting access to vital resources such as food, clean water, and healthcare~\citep{WorldBank2016}. These deprivations are expected to raise both mortality and morbidity rates.
According to WHO projections, climate change could be responsible for 250,000 additional deaths annually between 2030 and 2050, primarily through direct health threats~\citep{world2014quantitative}.

Furthermore, climate change could increase global mortality risk by approximately 3.2\% of global GDP by 2100 under a high emissions scenario according to projections by~\cite{carleton2022valuing}.\footnote{Damages are expressed in terms of the Value of a Statistical Life (VSL), an economic measure of the value placed on reducing the risk of death. The VSL is derived from the willingness to pay for marginal reductions in mortality risk. It is used in cost-benefit analyses to assess the economic impacts of policies that affect health and safety.}

Beyond macroeconomic estimates, climate change is most notably expected to reshape individual health trajectories and treatment needs, altering disease incidence and drug demand, particularly in the context of rising chronic conditions and increased exposure to environmental stressors.
Due to global warming, the Northern Hemisphere is expected to face not only hotter summers but also more extreme winter conditions. Although average winter temperatures may rise, climate projections indicate an increased likelihood of severe winter storms. These events often bring strong winds that amplify cold exposure, particularly among vulnerable populations, thereby increasing the risk of cold-related conditions such as hypothermia and frostbite. In addition, prolonged cold stress can impair immune defenses, making individuals more susceptible to respiratory infections~\citep{redshaw2013potential}.

Rising temperatures, related increased air pollution and extended pollen seasons intensify respiratory and cardiovascular conditions~\citep{tran2023climate,d2014climate,khraishah2022climate}. In parallel, water and soil quality degradation contribute to a rise in diarrheal diseases, especially in vulnerable populations. Furthermore, the spread of vector-borne diseases is extending into non-subtropical regions~\citep{rocklov2020climate}, and rising anxiety, depression, and stress (related to extreme weather events, displacement, and environmental degradation) are increasingly associated with climate-related stressors and environmental disruptions~\citep{cianconi2020impact}. 
These diverse and compounding health stressors are expected to translate into increased pharmaceutical needs across populations. The multifaceted impacts of climate on both physical and mental health are expected to increase the consumption of medications for both targeted treatments (e.g., antiprotozoals, psychotropics) and general symptom management (e.g., analgesics)~\citep{redshaw2013potential}. Yet, the relationship between environmental stressors and pharmaceutical demand remains underexplored, even though pharmaceutical consumption could serve as a valuable proxy for climate-related health impacts.

Climate change generally refers to long-term transformations in climate systems, but its health impacts can materialize through more immediate and localized events. This study focuses on short- to medium-term fluctuations in environmental conditions, which can be considered a tangible expression of climate change. In this sense, increased medication use may reflect heightened exposure to climate-related stressors, providing a quantifiable signal of population-level health impacts. 
Understanding these dynamics is crucial to anticipate healthcare needs and support pharmaceutical system resilience. Forecasting models can inform preparedness strategies, optimize resource allocation, and improve supply chain performance, particularly during climate-induced demand surges~\citep{tokovic2023impact}. Integrated forecasting models incorporating climate, epidemiological, and pharmaceutical data are essential to support policy responses and ensure equitable access to care. These complex interactions highlight the need for integrated surveillance systems, adaptation strategies, and mitigation efforts that can deliver both environmental and immediate health benefits ~\citep{haines2006climate}. Accurate forecasting enables a proactive strategy to identify vulnerabilities and enhance the robustness of pharmaceutical distribution, e.g., through stockpiling or diversification of high-demand pharmaceuticals. In this context, the present study adopts a quantitative, data-driven approach to forecast respiratory medication demand under climate variability, with the aim of informing both public health policy and pharmaceutical supply chain design.

Supply chain resilience can be enhanced by anticipating consumption trends in response to climate change and optimizing resource allocation and asset management, particularly during crises such as the COVID-19 pandemic. Nonetheless, assessing the impact of climate change on medical needs remains complex, as healthcare systems respond through a mix of short-term interventions and long-term policy adaptations—mechanisms that are often multifactorial, delayed, and mediated by socioeconomic conditions, infrastructure constraints, and limited access to granular health data ~\citep{rocque2021health}.
Against this backdrop, this study seeks to quantify the impact of climate change on pharmaceutical demand, with a specific focus on respiratory treatments.

Among the various health domains, respiratory diseases are especially climate-sensitive. According to~\cite{d2016climate}, global warming accentuates temperature variability, leading to higher concentrations of pollutants and extended pollen seasons, which increase allergen exposure and result in more frequent and severe asthma symptoms. Similarly, a survey conducted among physicians from the National Medical Association revealed that patients with COPD and asthma have experienced increased symptoms due to weather changes~\citep{sarfaty2014survey}. Consequently, the demand for asthma-related medications (e.g., inhalers and corticosteroids) and COPD treatments (e.g., bronchodilators and anti-inflammatory agents) is expected to increase during heightened environmental stress. Furthermore, projections from the system dynamics model proposed by~\cite{abir2025impact} suggest that the demand for the asthma pharmaceutical albuterol will likely increase across most age groups due to a heightened prevalence of asthma driven by climatic conditions.

Moreover,~\cite{eguiluz2020need} discuss the contribution of climate change in exacerbating respiratory diseases, particularly among children, by highlighting how rising temperatures and altered weather patterns intensify allergen exposure and air pollution levels, ultimately increasing the incidence and severity of conditions such as asthma and allergic rhinitis.
~\cite{burte2018association} report that climate change extends pollen seasons and increases allergen exposure, contributing to the higher incidence and severity of allergic rhinitis. \cite{anderegg2021anthropogenic} point out that pollen seasons across North America have been lengthening and intensifying due to anthropogenic climate change, likely driving higher demand for antihistamines, nasal sprays, and other allergy medications.

These findings indicate the need for more efficient resource allocation and supply chain management within the pharmaceutical industry. 
Therefore, health professionals are called to advocate for effective adaptation strategies to optimize the availability of medications~\citep{deng2020climate}.
In fact, while direct health impacts, such as mortality, increase due to climate change, adaptation measures, including improved healthcare access and better pharmaceutical supply chains, can mitigate these effects~\citep{abir2025impact}. 


This study focuses on Greece as a case study, leveraging weekly data over a seven-and-a-half-year period to explore the relationship between climate variability and respiratory pharmaceutical demand. First, the potential causal links between environmental variables and pharmaceutical consumption are investigated using Granger causality spectra, which allow us to examine frequency-specific interactions beyond standard time-domain approaches. Based on these insights, multiple forecasting models are then developed and compared —including Prophet, a Vector Autoregressive model with exogenous variables (VARX), Random Forest with Moving Block Bootstrap (MBB-RF), and Long Short-Term Memory (LSTM) neural networks—to evaluate the predictive power of climate variables and identify the most effective strategies for anticipating demand fluctuations. The results aim to support public health preparedness and pharmaceutical supply chain resilience under changing climatic conditions.

\section{Data}
Weekly pharmaceutical sales and climate data for Greece were used to develop time series models and machine learning algorithms, assessing climate-health relationships and improving future healthcare predictions.
\textit{Alira Health}\footnote{Alira Health, \url{https://www.alirahealth.com/}} provided data on sales of respiratory pharmaceuticals in Greece from January 4, 2016, to June 26, 2023. Pharmaceutical sales data were obtained through Sell-Out Data Acquisition, enabled by software integration with the Alira Health panel of 1,200 pharmacies evenly distributed across Greece. This procedure ensures data quality and reliability while removing all personally identifiable information in full compliance with the General Data Protection Regulation (GDPR). The statistical expansion was performed weekly, with geographic granularity across twenty distinct territories. It relied on a dynamic panel of pharmacies and incorporated socio-demographic factors to enhance the accuracy of the estimates. 

Respiratory drugs classified at the ATC1\footnote{The Anatomical Therapeutic Chemical (ATC) classification system is maintained by the WHO Collaborating Centre: \url{https://www.whocc.no/atc_ddd_index/}. Level 1 refers to the anatomical main group, e.g., R for the respiratory system.} level—encompassing the full spectrum of medications targeting the respiratory system—were used for model estimation and training. This choice was made to account for potential overlaps in therapeutic applications across different respiratory conditions. Following expert advice from Alira, who are highly familiar with the dataset, only prescription (Rx) medications were considered, as they are subject to stricter regulatory control and monitoring, thus offering more reliable data for analysis. This aggregated approach does not preclude the possibility of future investigations into specific therapeutic classes or active ingredients. Throughout the manuscript, the weekly time series of respiratory prescription pharmaceuticals sold in Greece will be referred to as \textit{drug demand} to streamline the discussion.

The statistical properties of the series were evaluated using the Augmented Dickey–Fuller test with lag order selected by the Bayesian Information Criterion (BIC). The test rejected the null hypothesis of a unit root under the specification with drift, indicating that the process is stationary around a non-zero mean. Further analysis of the univariate time series of respiratory drug demand is reported in the Supplementary material, Section~S3.

\begin{figure}[t!]
\centerline{\includegraphics[width=0.7\textwidth]{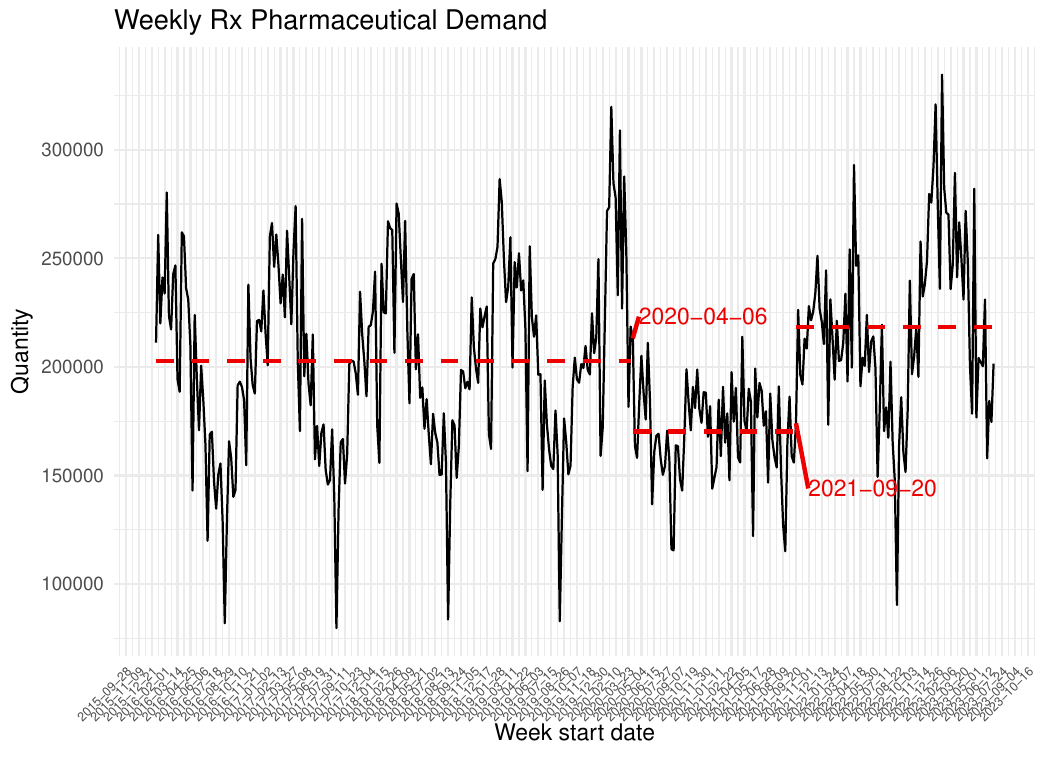}}
\caption{\textit{Weekly time series of prescription drug demand for respiratory system drugs (ATC1 R) in Greece, from January 4, 2016, to June 26, 2023. The red dashed lines indicate the mean levels within each segment defined by structural breaks, while the red week-start dates mark the breakpoints themselves, identified on April 6, 2020, and September 20, 2021. These structural changes were detected using the Bai–Perron multiple breakpoint algorithm, with the optimal number of breaks selected according to the BIC.}\label{fig:tot.qty_plot}}
\end{figure}

The drug demand time series exhibits structural breaks corresponding to the onset and the resolution of the COVID-19 pandemic emergency (see Figure~\ref{fig:tot.qty_plot}). These breakpoints were statistically identified using the~\cite{bai2003computation} procedure, which enables consistent estimation of multiple unknown breakpoints within a linear framework. Although the demand for ventilators and face masks increased substantially during this period, posing major challenges for the pharmaceutical and medical supply industries~\citep{ranney2020critical}, public demand for respiratory drugs decreased significantly. This decline was largely due to mitigation measures, such as lockdowns and mask-wearing, which reduced exposure to environmental, infectious, and allergenic triggers of respiratory diseases, ultimately resulting in structural shifts in pharmaceutical consumption patterns.

The climate data, primarily sourced from the ERA5 \footnote{\url{https://cds.climate.copernicus.eu/datasets/reanalysis-era5-single-levels}} reanalysis dataset~\citep{hersbach2020era5}, are available at hourly resolution. Data were collected at the geographical coordinates of the most populous city in each of the twenty pharmaceutical regions of Greece. A regional map showing the locations used for climate data extraction is provided in the Supplementary material, Section~S1. The extracted variables include the 2-meter air temperature ($Temp$), provided in Kelvin and converted to Celsius; the eastward and northward components of the wind at 10 meters above the surface of the Earth ($u_{10}$ and $v_{10}$, respectively, expressed in meters per second); total precipitation ($P$), representing accumulated liquid and frozen water, measured in meters and converted to millimeters; specific humidity, defined as the mass of water vapour per kilogram of moist air; and proportion of cloud cover. 
Wind speed was included because of its influence on pollutant and allergen transport and dispersion. It is derived for each hour $h$ and each region $r$ as
\begin{equation}
        WindSpeed_{r,h} = \sqrt{u_{10, r,h}^2 + v_{10, r,h}^2}.
\end{equation}
The data were aggregated at the daily and weekly levels, and subsequently at the national level using the mean, except for precipitation, which was aggregated using the sum to account for the cumulative nature of rainfall over time. To capture short-term climate variability, the number of wet days, extreme rainfall, and temperature standard deviation were derived for each week $w$ and each region $r$ as additional variables from ERA5 data as
\begin{eqnarray}
    WetDays_{r,w} &=& \sum_{d=1}^{7} H(P_{r,d} - 1 \, \text{mm})\\
    ExtremeRainfall_{r, w} &=& \sum_{d=1}^{7} H(P_{r,d} - P_{99.9, r}) \times P_{r,d}\\
    TemperatureSD_{r,w} &=& \sqrt{\dfrac{1}{7} \sum_{d=1}^{7}(Temp_{r,d} - \overline {Temp}_{r,w})^2}
\end{eqnarray}
where $H$ is the Heaviside step function, $P_{99.9, r}$ is the 99.9th percentile of historical daily precipitation for region $r$, 1 mm is the threshold used to define wet days, and $\overline{Temp}_{r,w}$ is the regional weekly average temperature. To spatially aggregate these variables across regions, wet days and temperature standard deviation were averaged, as they reflect relative intensity or frequency, while extreme rainfall was summed to represent the total burden of extreme precipitation events across the country. 

In addition to the variables mentioned above, the Fire Weather Index (FWI), developed by the Canadian Forestry Service, is sourced from the Copernicus Climate Data Store~\citep{copernicus2025}. The FWI combines the Initial Spread Index (ISI), which reflects the potential for fire spread, and the Buildup Index (BUI), which indicates the amount of fuel available for combustion, thus providing an integrated measure of forest fire conditions, including both the likelihood of ignition and the potential intensity and spread of fire. This index reflects fire risk and associated smoke-related air pollution, both of which are known to exacerbate respiratory conditions and potentially increase medication demand, particularly during hot and dry periods. The FWI was aggregated at the daily and weekly levels, and spatially using the mean.

\section{Methods}

Considering the complex interdependent relationships among various climate factors and their subsequent impact on the demand for respiratory medications, an exploratory analysis was necessary to identify relevant variables and appropriate lag structures. Based on the insights gained from this preliminary investigation, a set of predictive models incorporating the selected features and lag configurations was then employed to forecast drug demand.

\subsection{Frequency-domain causality analysis and feature selection}

The exploratory analysis began with unconditional and conditional Granger causality spectra to detect temporal dependencies that might be missed by time-domain methods, especially in cyclical data. This approach was motivated by periodic patterns observed in the Auto-Correlation Function (ACF) of drug demand and key environmental variables such as temperature and the FWI, highlighting the relevance of frequency-domain analysis.

The bootstrap procedure proposed by \cite{farne2022bootstrap} was used to assess whether observed relationships across frequencies were statistically significant. This approach is particularly advantageous compared to the parametric alternative proposed by~\cite{breitung2006testing} because it exhibits increasing power, approaching one, as the process becomes non-stationary. This makes it especially appropriate given the structural changes observed in the time series under investigation.

In the framework introduced by \citet{farne2022bootstrap}, the unconditional Granger causality spectrum quantifies how past values of one series help predict another across different frequencies. Specifically, for a bivariate stationary process $(X_t, Y_t)^T$ of length $T$, the unconditional Granger causality spectrum of the effect variable $Y_t$ with respect to the cause variable $X_t$, at frequency $\omega$, is defined as:
\begin{equation}
    h_{X\rightarrow Y}(\omega) = \log\left( \frac{ h_{YY}(\omega)}{\sigma_2 |\tilde{{P}}_{YY}(\omega)|^2}\right),
    \label{Unconditional GC}
\end{equation}
where $h_{YY}(\omega)$ denotes the spectral density of $Y_t$ at frequency $\omega$, $\tilde{P}_{YY}(\omega)$ is the $(2,2)$ element of the transformed transfer function matrix obtained from the normalized VAR representation of the process, and $\sigma_2$ is the variance of the innovation in $Y_t$ after applying the linear transformation that orthogonalizes the innovations. A positive value of $h_{ X\rightarrow Y}(\omega)$ implies that past values of $X_t$ improve the linear predictability of $Y_t$ at frequency $\omega$, with the corresponding cycle length given by $\omega^{-1}$.

For each of the $N$ stationary bootstrap samples simulated from the observed time series $(X_t, Y_t)$ following \cite{politis1994stationary}, resampled versions $X_t^*$ and $Y_t^*$ are generated. A VAR model is estimated on each bootstrap sample using Seemingly Unrelated Regressions (SURE) \citep{zellner1962efficient}. The optimal model order is determined by minimizing the BIC, which balances model parsimony and goodness of fit. Then, the Granger causality measure $h_{Y^* \to X^*}(2\pi f_i)$ at Fourier frequencies $f_i = i T^{-1}$, for $i = 1, \dots, \left\lfloor T/2 \right\rfloor$, is computed for each bootstrap sample. The median of these causality measures across all bootstrap samples is used to find the $(1-\alpha)$ quantile of the bootstrap median distribution. At each frequency $f_i$, the observed causality is flagged as significant if it exceeds the corresponding quantile of the bootstrap distribution.

A conservative Bonferroni correction is applied to control for multiple testing across the entire frequency domain. In this framework, the significance level for each individual frequency is adjusted to $ 2\alpha F^{-1} $, where $ F $ represents the total number of frequencies under consideration. This adjustment controls the family-wise error rate, thereby ensuring valid inference when testing for Granger causality at multiple frequencies simultaneously.

While the unconditional approach reveals overall dependencies, the conditional Granger causality Spectrum extends the unconditional version (\ref{Unconditional GC}) by controlling for an additional influencing variable, enabling the isolation of the direct causal effect between drug demand and climate-related variables while accounting for a potential confounder $W_t$~\citep{geweke1984measures}:
\begin{equation}
    h_{X\rightarrow Y | W}(\omega) = \log\left( \frac{h_{\tilde{Y}\tilde{Y}}(\omega)}{\sigma_{YY}|Q_{YY}(\omega)|^2}\right),
\end{equation}
where $Q_{YY}(\omega)$ is the $(2,2)$ element of the transfer function matrix derived from the VAR representation of the residual processes obtained by projecting both $X_t$ and $Y_t$ onto $W_t$ and removing the fitted components. $\sigma_{YY}$ denotes the variance of the residuals from the second equation of the conditional VAR, where $Y_t$ is the dependent variable after controlling for $W_t$.
The bootstrap methodology in this case requires estimating a VAR model on $Y_t$ and $W_t$, allowing for the simulation of $(Y_t^*, W_t^*)$ via residual bootstrap. Subsequently, the series $X_t^*$ is simulated independently using a stationary bootstrap. Finally, the conditional Granger causality spectrum is computed on each simulated triplet following the same steps as in the unconditional procedure.

Building on the insights gained from the frequency-domain analysis, the most influential variables for forecasting demand for respiratory medications were identified through feature importance analysis of the Random Forest model with Moving Block Bootstrap (MBB-RF). In standard Random Forests, each decision tree is trained on a bootstrap sample drawn with replacement from the original dataset~\citep{breiman2001random}, enhancing robustness by reducing overfitting. However, this approach assumes independent observations, making it unsuitable for time series data, which typically exhibit temporal dependence.

This limitation can be overcome through the use of the Moving Block Bootstrap (MBB) approach, as defined by \cite{kunsch1989jackknife}, which samples consecutive blocks of data, preserving the temporal structure and ensuring that the inherent time dependencies are maintained. 
The resulting approach, here referred to as Moving Block Bootstrap Random Forest (MBB-RF), was previously considered by~\citet{goehry2023random} as an adaptation of the Random Forest methodology to time series data (see Algorithm~\ref{alg:rf-mbb}).
The feature importance metric adopted for feature selection is the impurity-based measure, which quantifies the average reduction in residual sum of squares across all splits in the forest where the given predictor is used.

\begin{algorithm}[h!]
\caption{MMB-RF: Random Forest Regression with Moving Block Bootstrap Aggregation}
\label{alg:rf-mbb}

\textbf{Input:} \\
\hspace{1em} Time series data $\mathcal{D} = \{(\mathbf{X}_t, y_t)\}_{t=1}^{T}$, where $\mathbf{X}_t$ denotes the vector of predictors at time $t$, including lagged values of both the target variable and the covariates. \\
\hspace{1em} $B$: Number of trees in the ensemble \\
\hspace{1em} $m$: Number of variables randomly selected at each split \\
\hspace{1em} $n_{\min}$: Minimum node size (stopping criterion)

\begin{algorithmic}[1]
\For{$b = 1$ \textbf{to} $B$}
    \State \textbf{Bootstrap:} Generate a resampled dataset $\mathcal{D}^{*(b)}$ as follows:
    \Statex \hspace{2em} \textnormal{(a) Define $T - \ell + 1$ overlapping blocks of length $\ell$:}
    \[
        \mathcal{B}_i = \{(\mathbf{X}_i, y_i), \dots, (\mathbf{X}_{i+\ell-1}, y_{i+\ell-1})\}, \qquad i = 1, \dots, T - \ell + 1
    \]
    \Statex \hspace{2em} \textnormal{(b) Draw $K = \lceil T / \ell \rceil$ blocks with replacement from $\{\mathcal{B}_i\}$}
    \Statex \hspace{2em} \textnormal{(c) Concatenate the sampled blocks in temporal order:}
    \[
        \mathcal{D}^{*(b)} = \text{concat}(\mathcal{B}_{i_1}, \dots, \mathcal{B}_{i_K})
    \]
    \Statex \hspace{2em} \textnormal{(d) Truncate $\mathcal{D}^{*(b)}$ to length $T$ if necessary}
    
    \State \textbf{Tree Growing:} Train regression tree $T_b$ on $\mathcal{D}^{*(b)}$
    \While{node size $> n_{\min}$ and variance $> 0$}
        \State Randomly select $m$ predictors
        \State Split the node using the feature and threshold that minimize the within-node variance
        \State Grow child nodes recursively
    \EndWhile
\EndFor

\vspace{1em}
\textbf{Prediction:} For a new observation $\mathbf{x}$,
\[
    \hat{y}_{\text{RF}}(\mathbf{x}) = \frac{1}{B} \sum_{b=1}^{B} T_b(\mathbf{x})
\]
\end{algorithmic}
\end{algorithm}

In parallel to the nonparametric analysis performed through the MBB-RF, a sparse VAR model was also estimated to gain a complementary understanding of the dynamic interrelationships between drug demand and environmental predictors. Specifically, we employed a penalized VAR model with LASSO (Least Absolute Shrinkage and Selection Operator) penalty~\citep{basu2015regularized}. This approach simultaneously enables variable selection and parameter estimation, making it well suited for high-dimensional time series with potentially correlated regressors. The use of a sparse VAR provides an interpretable framework in which only the most relevant temporal dependencies are retained, with many coefficients shrunk exactly to zero.

\subsection{Forecasting models}

Establishing a solid forecasting baseline began with implementing the Prophet model~\citep{taylor2018forecasting}, which was selected for its ability to capture complex temporal dynamics, including multiple seasonalities and structural breaks. This choice is particularly appropriate given that the time series of prescription pharmaceutical sales shows seasonal behavior and structural changes in the level.

The Prophet model, which relies solely on the target’s past values, offering a purely autoregressive benchmark of the target series, serves as a benchmark to assess whether the inclusion of climate covariates significantly improves forecast accuracy. Let $S$ be the number of changepoints and $\mathbf{a}_t$ a binary vector indicating whether time $t$ occurs after each changepoint $s_j$, for $j = 1, \dots, S$. Prophet decomposes the observed time series $y_t$ into an additive model including trend ($g_t$), seasonality ($s_t$), holidays ($h_t$), and a Gaussian error term $\epsilon_t \sim \mathcal{N}(0, \sigma^2)$:
\begin{equation}
    y_t = g_t + s_t + h_t + \epsilon_t.
\end{equation}

The trend component $g_t$ is modeled as a piecewise linear function:
\begin{equation}
    g_t = \left(k + \mathbf{a}_t^\top \mathbf{\delta}\right)t + \left(m + \mathbf{a}_t^\top \mathbf{\gamma}\right),
    \label{trend_prophet}
\end{equation}
where $k$ is the base growth rate and $\mathbf{\delta}$ adjusts the rate at each changepoint. The offset term includes the initial offset $m$ and a correction vector $\mathbf{\gamma}$, where each $\gamma_j = -s_j \delta_j$ ensures continuity at changepoints. Prophet automatically selects changepoints from a predefined grid by placing a sparse prior on $\mathbf{\delta}$.

The seasonal component $s_t$ is expressed using a truncated Fourier series   to  provide a flexible model of periodic effects~\citep{harvey199310}:
\begin{equation}
    s_t = \sum_{n=1}^{N} \left[ a_n \cos\left(\frac{2\pi n t}{P}\right) + b_n \sin\left(\frac{2\pi n t}{P}\right) \right],
\end{equation}
where $P$ denotes the period of seasonality and $N$ controls the number of harmonics. Finally, the holiday component $h_t$ captures the effect of known recurring events by introducing indicator variables for each holiday and estimating their impact with separate parameters~\citep{taylor2018forecasting}.

A VARX model, a MBB-RF, and a LSTM neural network were employed to incorporate climate-related variables into the prediction of drug demand. These models were selected to represent a diverse set of methodological approaches—linear, ensemble-based, and deep learning—each capable of capturing different aspects of the relationship between climatic conditions and respiratory drug consumption.

After verifying the absence of cointegration among the selected variables using the Johansen test, a VARX model with exogenous regressors was estimated.
Considering a set of $K$ endogenous variables and $M$ exogenous predictors, a Vector Autoregressive model ~\citep{lutkepohl2005new} with exogenous regressors (VARX) can be expressed as
\begin{equation}
\mathbf{y}_t = \mathbf{\nu} + \sum_{i=1}^{p} \mathbf{A}_i \mathbf{y}_{t-i} + \mathbf{B} \mathbf{x}_t + \mathbf{u}_t,
\label{eqn:VARX}
\end{equation}
where $\mathbf{y}_t$ is a $K$-dimensional vector of endogenous variables, $\mathbf{x}_t$ is an $M$-dimensional vector of time-aligned exogenous covariates, and $\mathbf{\nu}$ is a $K$-dimensional vector of intercepts. The matrices $\mathbf{A}_i$ ($K \times K$) capture the autoregressive structure over $p$ lags, while $\mathbf{B}$ ($K \times M$) quantifies the contemporaneous effects of the exogenous covariates. The innovation term $\mathbf{u}_t$ is assumed to be a white noise process with zero mean and constant covariance matrix. The optimal lag order $p$ is selected using the BIC.

Although the marginal distributions of the endogenous variables exhibit limited kurtosis, several display noticeable skewness or bimodality (see Supplementary Material, Section~S2). Furthermore, Shapiro–Wilk tests reject the null hypothesis of normality for most variables at the conventional 5\% significance level.

The VARX model was estimated using a residual bootstrap procedure~\citep{kilian1998small} to ensure valid inference in the presence of such deviations from normality. This approach does not rely on asymptotic distributional assumptions and yields more accurate confidence intervals and test statistics in finite samples (in this case, $T = 390$). Furthermore, it enhances the robustness of inference even when residuals appear approximately normal, as it accommodates nonstandard features in the marginal distributions of the input variables.


Centered residuals $\hat{\mathbf{u}}_t = \mathbf{u}_t - \bar{\mathbf{u}}$ from (\ref{eqn:VARX}) are sampled with replacement to obtain bootstrap residuals $ \mathbf{u}_t^* $, which are then used to generate bootstrap samples $ \mathbf y_t^{*(b)}$ recursively, for $b = 1, \dots, B$:
\begin{equation}
\mathbf{y}_t^{*(b)} = \hat{\mathbf{\nu}} + \sum_{i=1}^{p} \hat{\mathbf{A}}_i \mathbf{y}_{t-i}^{*(b)} + \hat{\mathbf{B}} \mathbf{x}_{t} + \mathbf{u}_t^{*(b)}, \quad t = p+1, \dots, T.
\end{equation}

For each simulated series $ \mathbf y_t^{*(b)}$, bootstrap estimates $\hat{\mathbf{A}}_i^{(b)}$ and $\hat{\mathbf{B}}^{(b)}$ are computed. Bias-corrected coefficients are then obtained as \footnote{If the corrected parameters $\tilde{\mathbf{A}}_i$ and $\tilde{\mathbf{B}}$ violate the stability condition (i.e., at least one eigenvalue of the companion matrix lies within or on the unit circle), the bias terms are progressively shrunk by a factor $\delta < 1$ until stability is restored.}
\begin{equation}
\tilde{\mathbf{A}}_i = \hat{\mathbf{A}}_i - \left( \frac{1}{B} \sum_{b=1}^{B} \hat{\mathbf{A}}_i^{(b)} - \hat{\mathbf{A}}_i \right), \qquad
\tilde{\mathbf{B}} = \hat{\mathbf{B}} - \left( \frac{1}{B} \sum_{b=1}^{B} \hat{\mathbf{B}}^{(b)} - \hat{\mathbf{B}} \right).
\end{equation}

The MBB-RF approach was employed as a machine learning forecasting method due to its ability to capture complex nonlinear relationships and interactions among predictors without requiring strong parametric assumptions. The training procedure adopted for forecasting mirrors the one used during the variable selection presented in the previous subsection (Algorithm~\ref{alg:rf-mbb}), ensuring methodological consistency in the modeling of temporal dependencies and lagged structures.

A Long Short-Term Memory (LSTM) neural network~\citep{hochreiter1997long} was implemented to assess whether a deep learning model offers performance gains over traditional approaches in capturing the potentially nonlinear and complex relationship between climatic factors and drug demand. LSTMs are specifically designed for sequential data and are capable of learning long-range dependencies through memory cells that maintain internal states over time. At each time step $t = 1, \dots, T$, the state $s_{c_j}(t)$ of memory cell $c_j$ evolves through the interaction of input and output gates, which regulate the flow of information. This gating mechanism allows the network to retain or discard information selectively, thus effectively mitigating the vanishing gradient problem that affects standard Recurrent Neural Networks (RNNs) and limiting their ability to model long-term dependencies~\citep{hochreiter1991untersuchungen}.

The update rule for the internal state is:
\begin{equation}
s_{c_j}(t) = s_{c_j}(t-1) + y^{in_j}(t) \cdot g\left(\sum_u w_{c_j,u}y^u(t-1)\right), \qquad s_{c_j}(0) = 0,
\end{equation}
where $ g(\cdot) $ is a nonlinear activation function that controls how new information is incorporated into the memory and $ y^u(t-1) $ denotes the output of unit $ u $ at the previous time step. The input and output gate activations are defined as:
\begin{equation}
y^{in_j}(t) = f_{in_j}\left(\sum_u w_{in_j,u} \cdot y^u(t-1)\right); \qquad
y^{out_j}(t) = f_{out_j}\left(\sum_u w_{out_j,u} \cdot y^u(t-1)\right),
\end{equation}
where $ w_{in_j,u}, w_{out_j,u} \in \mathbb{R} $ are trainable connection weights determining the influence of input $ u $ on the respective gate unit $ j $ and $f_{in_j}$ and $f_{out_j}$ are activation functions that control the openness of the input and output gates. The output of the memory cell is then given by:
\begin{equation}
y^{c_j}(t) = y^{out_j}(t) \cdot h(s_{c_j}(t)),
\end{equation}
where $ h(\cdot) $ is an output activation function applied to the internal state.


SHapley Additive exPlanations (SHAP) values were employed to enhance the LSTM model interpretability and elucidate the contribution of individual predictors. SHAP is grounded in cooperative game theory and assigns each feature an importance score corresponding to its average marginal contribution across all possible subsets of features~\citep{lundberg2017unified}. This yields explanations that are both locally accurate and globally consistent, effectively integrating the strengths of global attribution methods, such as Permutation Importance~\citep{breiman2001random}, and local interpretability tools like LIME~\citep{ribeiro2016why}. Differently from Permutation Importance, SHAP is robust to multicollinearity, as it evaluates conditional expectations rather than marginal perturbations. Moreover, it avoids the instability and inconsistency that may affect LIME’s local surrogate models. Despite its computational cost, SHAP’s additive nature and sound theoretical grounding make it especially well-suited for interpreting complex, nonlinear architectures such as LSTMs.

\subsection{Performance evaluation criteria}

Forecasting models were trained on the first 338 weekly observations (approximately six and a half years) and evaluated on a hold-out test set consisting of the final 52 weeks. A set of widely used error metrics was computed on the test set to evaluate forecasting performance, including the Mean Absolute Percentage Error (MAPE), Root Mean Squared Error (RMSE), RMSE-observations Standard deviation Ratio (RSR), and the seasonal Mean Absolute Scaled Error (MASE)~\citep{hyndman2006another,hyndman2018forecasting,moriasi2007model}. Let $ y_h $ denote the observed value at horizon step $ h $, and $ \hat{y}_h $ the corresponding predicted value, for $ h = 1, \dots, H $, where $ H = 52 $ is the test set length. The error metrics are formally defined as:
\begin{equation}
\begin{array}{ll}
\text{MAPE} = \dfrac{100}{H} \displaystyle \sum_{h=1}^{H} \left| \dfrac{y_h - \hat{y}_h}{y_h} \right|, &
\qquad \text{RMSE} = \sqrt{ \dfrac{1}{H} \displaystyle \sum_{h=1}^{H} (y_h - \hat{y}_h)^2 },\\[10pt]
\text{RSR} = \sqrt{\dfrac{ \displaystyle \sum_{h=1}^{H} (y_h - \hat{y}_h)^2 }{ \displaystyle \sum_{h=1}^{H} (y_h - \bar{y}_{\text{train}})^2 }},&
\qquad \text{MASE} = 
\frac{ \frac{1}{H} \displaystyle \sum_{h=1}^{H} \left| y_h - \hat{y}_h \right| }
{ \frac{1}{T - m} \displaystyle \sum_{h=m+1}^{T} \left| y_h - y_{t - m} \right| }.
\end{array}
\end{equation}

Here, $T$ is the length of the training set, $\bar{y}_{\text{train}}$ denotes the mean of the observed values in the training set, and $m$ represents the seasonal lag used for scaling in the MASE metric, which is set to $m = 52$ to reflect the annual seasonality inherent in weekly data.

The use of multiple evaluation metrics enables a more nuanced understanding of model performance, capturing scale-independent errors, sensitivity to outliers, and relevance to domain-specific seasonal patterns. Specifically, MAPE assigns equal percentage weight to each observation, making it useful for evaluating relative errors across scales. RMSE quantifies the magnitude of typical prediction errors, placing greater emphasis on large deviations due to the squaring of residuals. Notably, RSR is mathematically related to $R^2$ via the identity $R^2 = 1 - \text{RSR}^2$, allowing for a complementary interpretation. RSR provides a normalized error version of RMSE, facilitating comparisons across series with different levels of variability, while $R^2$ measures the proportion of variance in the observed series explained by the model, making it particularly useful for assessing overall model fit. Finally, the seasonal version of MASE evaluates forecast accuracy relative to a naïve yearly benchmark, making it particularly suitable for time series with recurring seasonal patterns.

Each metric was calculated individually for the forecasting models, allowing for a detailed assessment of predictive accuracy and a direct comparison of their effectiveness in forecasting drug demand. This evaluation also offered insight into the added value of incorporating climate-related variables by benchmarking each model against a univariate Prophet baseline relying solely on the autoregressive dynamics of the target series.

It is important to note that among the implemented models, the VARX model is distinct in its forecasting strategy, as it operates through a recursive one-step-ahead approach, in which each prediction is iteratively fed back into the model to generate the subsequent forecast. In contrast, the Prophet, MBB-RF, and LSTM models are trained to produce the entire 52-week forecast horizon in a single step. This distinction is particularly relevant as recursive approaches in autoregressive frameworks are prone to error accumulation, particularly over longer horizons. Conversely, direct forecasting strategies mitigate the propagation of errors but may require larger training datasets to achieve reliable performance.

This comprehensive evaluation framework ensures that model comparisons are both methodologically rigorous and practically informative, particularly in guiding policy decisions related to respiratory health planning.

\subsection{Software and implementation}





Forecasting models such as Prophet, VAR, and MBB-RF were implemented in \texttt{R} (version~4.4.3) using the \texttt{prophet}\citep{prophetR}, \texttt{vars}\citep{pfaff2018package}, and \texttt{rangerts}\citep{huiyan2020rangerts} packages, respectively. The LSTM neural network was developed in \texttt{Python} (version~3.11.9) with the \texttt{TensorFlow} and \texttt{Keras} libraries. Data preprocessing, visualization, and performance evaluation were primarily conducted in \texttt{R}, complemented by \texttt{pandas}, \texttt{matplotlib}, and \texttt{numpy} for auxiliary tasks in Python.

Frequency-domain Granger causality tests, including the conditional version, were carried out in \texttt{R} via the \texttt{grangers}\citep{farne2018r} package, which implements the methodology of Farnè and Montanari (2023). Additionally, the \texttt{sparsevar} package\citep{vazzole2023sparsevar} was employed to explore sparse VAR representations and dynamic interactions in high-dimensional settings.

Interpretability of the LSTM model was addressed in \texttt{Python} using the \texttt{shap} library to compute Shapley values and evaluate feature contributions. Predictive uncertainty was assessed through Monte Carlo dropout by generating multiple stochastic forward passes during inference.

All code used for model development and analysis is available upon request and will be made publicly accessible upon publication to promote transparency and reproducibility.

\section{Results}
\subsection{Granger Causality Spectra}
\begin{figure}[t!]
\centering
\includegraphics[width = \textwidth]{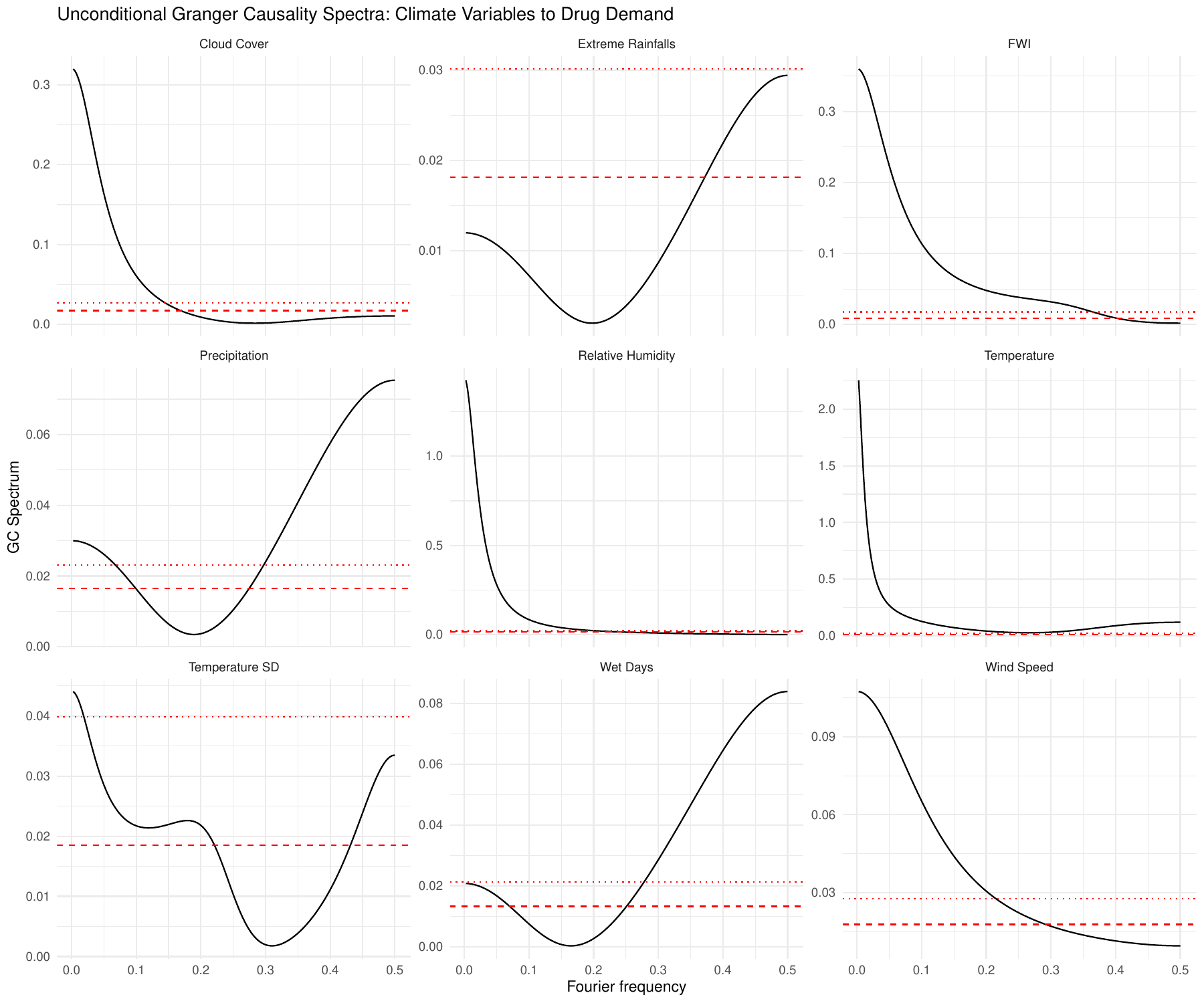}
\caption{\textit{Unconditional Granger causality spectra from climatic variables to drug demand. Each panel displays the frequency-specific Granger causality spectrum from a given predictor to the weekly quantity of respiratory drug prescriptions. The solid line represents the estimated spectrum. The dashed line denotes the 95\% significance threshold under the null hypothesis of no Granger causality, based on the empirical distribution of the bootstrap test statistic. The dotted line corresponds to the Bonferroni-adjusted threshold that controls for multiple testing across frequencies.}
\label{fig:gc_uncond}}
\end{figure}

\begin{figure}[t!]
\centering
\includegraphics[width = \textwidth]{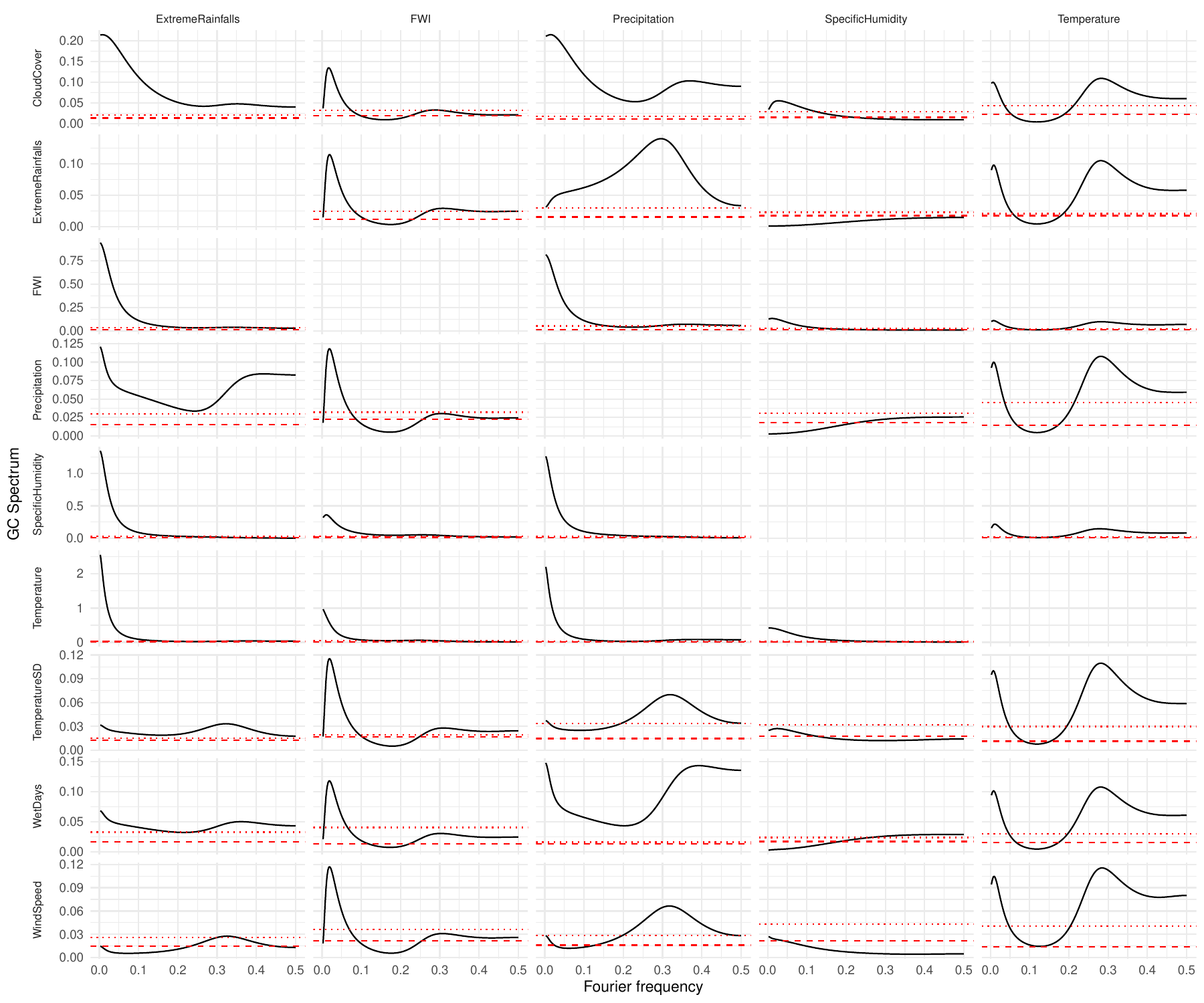}
\caption{\textit{Conditional Granger causality spectra (part 1). Each panel shows the frequency-specific conditional Granger causality from a climatic variable $X_t$ in row to drug demand, while controlling for a third variable $W_t$ in column. The solid line represents the estimated spectrum. The dashed line indicates the 95\% significance threshold derived from bootstrap inference under the null hypothesis of no conditional Granger causality. The dotted line shows the Bonferroni-adjusted threshold, accounting for multiple testing across frequencies.}
\label{fig:gc_cond1}}
\end{figure}

\begin{figure}[t!]
\centering
\includegraphics[width = \textwidth]{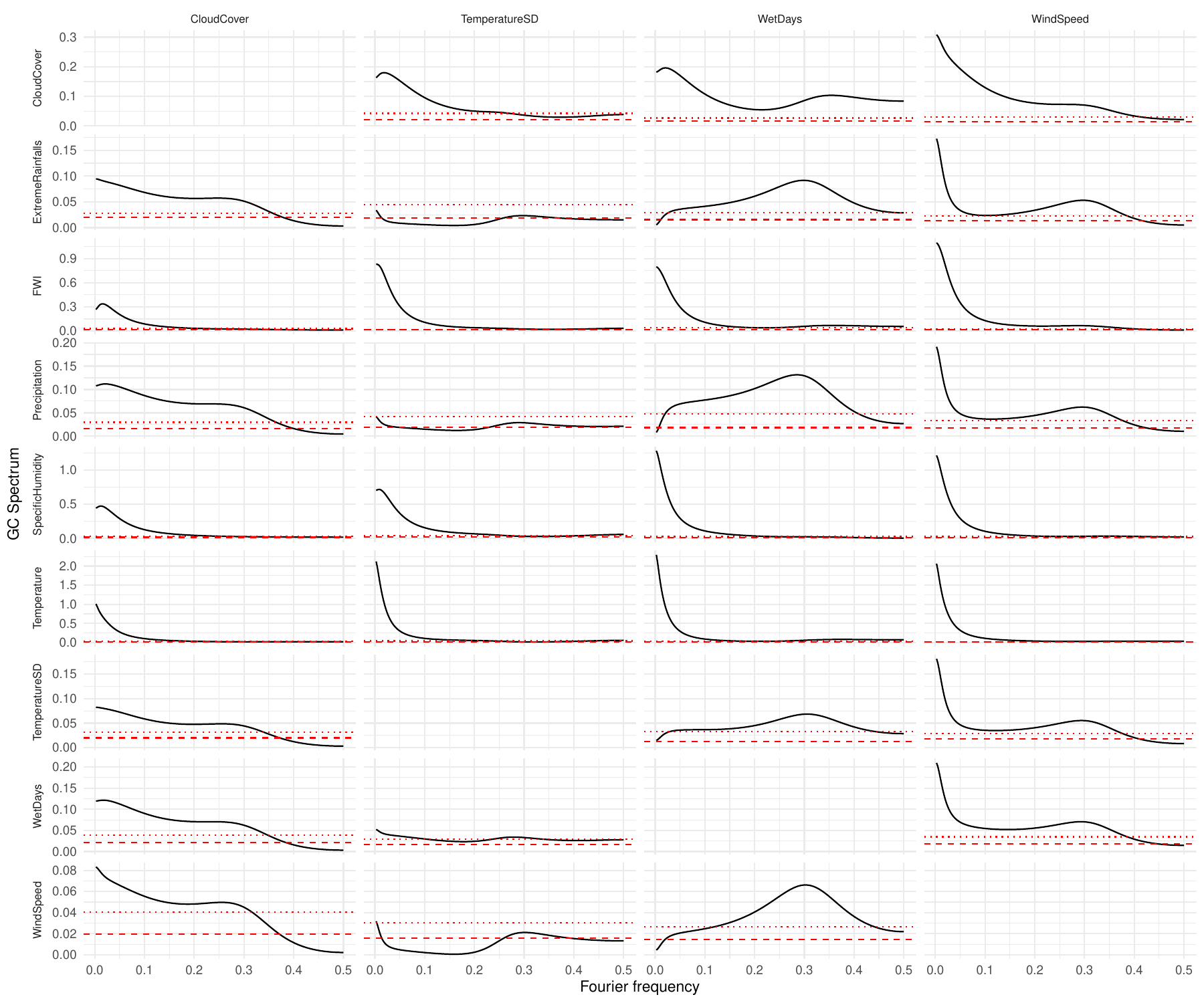}
\caption{\textit{Conditional Granger causality spectra (part 2). Each panel shows the frequency-specific conditional Granger causality from a climatic variable $X_t$ in row to drug demand, while controlling for a third variable $W_t$ in column. The solid line represents the estimated spectrum. The dashed line indicates the 95\% significance threshold derived from bootstrap inference under the null hypothesis of no conditional Granger causality. The dotted line shows the Bonferroni-adjusted threshold, accounting for multiple testing across frequencies.}
\label{fig:gc_cond2}}
\end{figure}

The analysis of both unconditional and conditional Granger causality spectra (Figures~\ref{fig:gc_uncond},~\ref{fig:gc_cond1}, and~\ref{fig:gc_cond2}) reveals the frequency-dependent influence of climatic variables on drug demand. To enhance the characterization of the temporal structure of these relationships, we extracted the cyclical component of each series using the Hodrick–Prescott (HP) filter~\citep{hodrick1997postwar}. Following the adjustment proposed by~\cite{ravn2002adjusting} for higher-frequency data, the smoothing parameter was set to $\lambda = 1600 \times (52/4)^4$ to account for weekly observations. This specification allows for the removal of long-term trends while preserving short- and medium-term fluctuations that are more likely to reflect meaningful causal interactions.

Unconditional Granger causality spectra (Figure~\ref{fig:gc_uncond}) reveal that all examined climate variables exhibit statistically significant causal effects, indicating that multiple environmental factors contribute to fluctuations in respiratory drug consumption. Notably, temperature, specific humidity, wind speed, and the FWI display spectra that peak at low frequencies and gradually decline while remaining statistically significant at medium frequencies. Such associations exceed both the standard significance threshold and the more conservative Bonferroni-adjusted threshold, underscoring the robustness of their predictive value. The observed spectral pattern—significant at both low and medium frequencies—indicates that long-term seasonal trends and mid-range climatic fluctuations both contribute to shaping respiratory drug demand.

Cloud cover shows a similar peak at low frequencies, but its spectrum decreases more steeply and falls below significance thresholds at medium frequencies, indicating a less persistent influence. In contrast, precipitation and the number of wet days show significant spectral peaks at higher frequencies, pointing to short-term, event-driven effects likely associated with acute weather phenomena. Meanwhile, the spectra for extreme rainfall and temperature standard deviation are comparatively weak, with significant peaks restricted to a narrow range of frequencies—high for extreme rainfall and low for temperature variability. These patterns suggest that, although extreme rainfall and temperature variability may have some localized, short-lived impacts, they are not consistent or dominant drivers of respiratory drug demand in the longer term. This highlights the importance of distinguishing between transient weather anomalies and systematic climatic shifts, especially in the context of climate change, which is expected to increase both the frequency and intensity of extreme events.

Conditional Granger causality spectra (Figures~\ref{fig:gc_cond1} and~\ref{fig:gc_cond2}) offer deeper insights into the distinct role of each climate variable by controlling for the confounding influence of the others, enabling the disentanglement of direct from indirect effects and assess the unique contribution of each variable to the prediction of drug demand. 

Consistent with the unconditional results, temperature, specific humidity, wind speed, and the FWI remain among the most robust predictors, though the signal for wind speed appears comparatively weaker. Their conditional spectra frequently exhibit significant peaks, especially at low frequencies, even after adjusting for other variables. This pattern suggests that part of their predictive power is shared with other climatic factors, yet their influence is not fully mediated, reinforcing their direct relevance for respiratory drug demand.

Certain interactions between variables reveal more intricate dynamics. Notably, using precipitation as a conditioning variable enhances the spectral intensity of otherwise moderate predictors such as wet days, cloud cover, and extreme rainfall. Likewise, the spectrum of precipitation itself becomes more pronounced when conditioned on wet days or extreme rainfall. These mutual amplifications likely reflect shared meteorological processes—such as the temporal clustering and persistence of precipitation events—that emerge more clearly once redundant variability is partialled out.

By contrast, specific humidity as a conditioning variable tends to flatten most spectra, substantially reducing the apparent influence of many predictors. Even the spectrum of temperature, typically a dominant climatic driver, becomes less pronounced. While still statistically significant, it shows a flatter shape, with a subtle and low peak at low frequencies. This likely reflects the strong correlation between humidity and temperature, particularly in their seasonal behavior, suggesting that humidity absorbs a considerable portion of the shared explanatory power when both are included.

The behavior of FWI is also informative. Its conditional spectra remain strong across most combinations, reaffirming the robustness observed in the unconditional analysis. However, its predictive signal weakens considerably when temperature or humidity are used as conditioning variables, indicating that a substantial part of FWI predictive power may be mediated through these two core drivers of atmospheric and fire-related dynamics. This interpretation aligns with the known formulation of FWI, which includes components such as air temperature, specific humidity, and wind speed.

Interestingly, temperature and FWI as conditioning variables enhance the spectral signatures of several weaker predictors, including wet days, cloud cover, and temperature standard deviation. These patterns suggest that once the dominant seasonal component of temperature or fire-weather-related variability is controlled for, subtler dynamics in the remaining variables emerge. Nonetheless, temperature and humidity as causal variables lose spectral strength when FWI is the conditioning variable, again indicating a high degree of shared variance likely tied to seasonal cycles.

Overall, these findings underscore the importance of addressing multicollinearity and indirect causal pathways in climate–health models. They also highlight that even predictors with weak unconditional effects may become relevant under specific conditional scenarios, reinforcing the utility of frequency-domain Granger causality in revealing hidden structures in environmental influences on health outcomes.

\subsection{Variable selection}
The role of climate-related variables in predicting drug demand is further assessed through a variable importance analysis conducted using the MBB-RF model, as described in Algorithm~\ref{alg:rf-mbb}. Since random forests were not originally designed for time series applications, it is necessary to define the appropriate number of lags to include in the feature set prior to model training. The optimal lag order was identified using the BIC applied to VAR models estimated for both unconditional and conditional Granger causality testing. Since no model selected an order greater than four, this lag structure was uniformly adopted across all variables during the training phase. The weekly resolution of the data further supports this choice, as a lag of four weeks approximately corresponds to one month, aligning with common pharmaceutical purchasing behaviors. Indeed, medications prescribed for respiratory conditions are typically issued monthly due to clinical factors—such as standard dosing regimens and regular follow-up requirements—and administrative constraints, including limits on repeat prescription durations and the implementation of monthly therapeutic plans. Also, setting the number of lags to four satisfies a principle of model parsimony, limiting complexity without sacrificing interpretability.

Each series was assessed for weak stationarity to validate the MBB approach. The RF model was trained using 1,000 trees, each built on an MBB sample with a block size of 52 weeks, reflecting the seasonal nature of the data. Since respiratory drug consumption follows a clear annual cycle (see Figure~S5 in Supplementary material), this block size preserves the intra-annual temporal dependencies critical to model performance. Each tree was grown using a minimum node size of five, following standard recommendations in machine learning literature for regression tasks~\citep{wright2017ranger, breiman2001random}. This setting balances model flexibility and overfitting control, especially in time series contexts where deeper trees may lead to instability due to autocorrelated inputs.

Given the length of the training set ($T = 338$), this setup yields 287 overlapping blocks of 52 weeks each. For each tree, 7 blocks were drawn with replacement, resulting in a training sample of 364 observations per tree. By concatenating the selected blocks, the resampled series retains its temporal coherence within each block, ensuring that key seasonal patterns are not disrupted during the bootstrap resampling process.

\begin{figure}[h]
\centering
\includegraphics[scale=0.45]{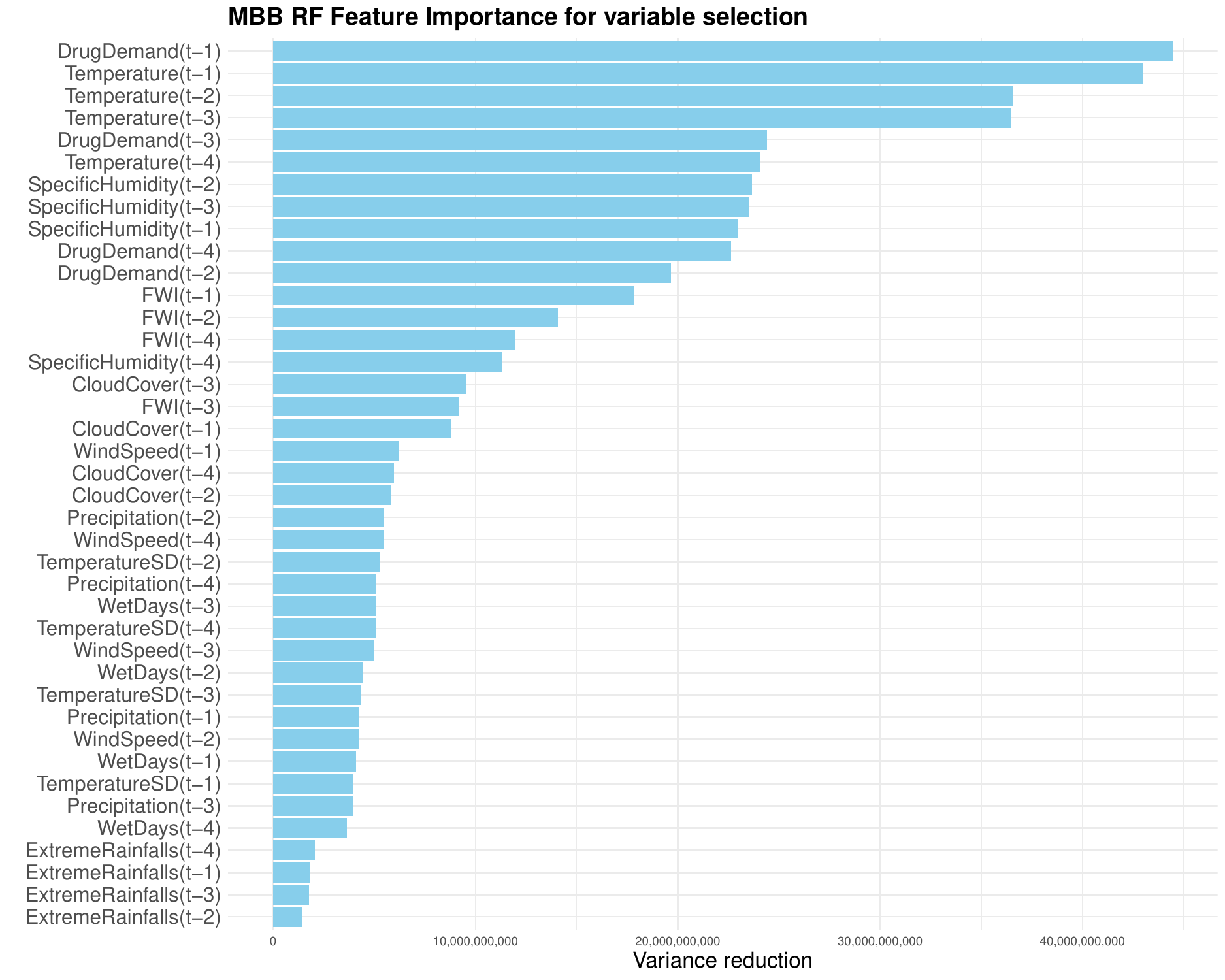}
\caption{\textit{Feature importance scores computed during the variable selection phase using a Random Forest model with Moving Block Bootstrap (MBB-RF). Importance is quantified as total impurity reduction, averaged over bootstrap replicates. The plot highlights the most informative predictors for inclusion in the forecasting models.}}
\label{fig:rf_importance}
\end{figure}

The results from the Granger causality analysis were integrated to the variable importance rankings obtained from the MBB-RF procedure (Figure~\ref{fig:rf_importance}) to guide the selection of predictors for the forecasting models.

Given the strong autoregressive nature of the drug demand series, lagged values of the target variable were included as predictors in all forecasting models. Indeed, autoregressive terms emerge among the most important features, highlighting the persistence and temporal structure of respiratory drug consumption. The resulting importance plot reveals a pronounced elbow shape, suggesting that only a small subset of variables substantially contributes to predictive accuracy. This curvature is used to define a threshold for variable selection, effectively identifying the most informative predictors while discarding less relevant ones. The lagged drug demand variables dominate the top ranks along with temperature lags. In contrast, extreme rainfall and temperature standard deviation appear among the least relevant predictors, which is in line with their limited significance in both unconditional and conditional Granger causality spectra.

Among the climate variables, temperature and specific humidity consistently rank highly. However, their strong static correlation (0.9449) indicates substantial redundancy and raises concerns about multicollinearity, which is particularly problematic for VAR models. While MBB-RF and LSTM tend to be more robust to collinearity, the same predictor set was initially applied across all models to ensure comparability of forecasting performance. Despite the relevance of specific humidity, supported by both Granger causality and feature importance analyses, its inclusion alongside temperature introduced overlap without leading to consistent improvements in accuracy. Therefore, specific humidity was excluded in favor of temperature, which empirically emerged as the more informative and impactful predictor.

This selection strategy, guided by the elbow in the importance curve, led to a parsimonious yet effective predictor set, composed solely of temperature and the lagged values of respiratory drug demand, capturing both the intrinsic temporal dynamics of the target series and the influence of the most relevant climate-related driver.

A sparse VAR(4) model with a LASSO penalty (i.e., $\ell_1$ regularization) was estimated to further assess the system’s predictive structure. Consistent with the MBB-RF model, the lag order was set to $p = 4$ to capture monthly cyclic dynamics, as suggested by the autocorrelation functions. All variables were standardized to ensure a fair regularization process, preventing dominance by high-variance predictors.

The resulting model is highly interpretable due to the sparsity induced by LASSO. In the equation for drug demand (Table~\ref{tab:lasso}), the response exhibits strong autoregressive behavior, with a dominant lag-1 coefficient ($0.2575$) and smaller positive effects at lags 2–4. This confirms the temporal persistence supported by both the Granger spectral analysis and the MBB-RF model.

Among climatic variables, temperature displays consistently negative coefficients at lags 1 and 3, suggesting that warmer weather reduces respiratory drug demand—likely due to lower infection transmission and improved air quality. This interpretation aligns with the importance scores from the MBB-RF model, where temperature emerged as a key predictor.

Other variables, such as wind speed and precipitation, show smaller and more delayed effects. Cloud cover appears to have limited influence. Specific humidity was retained in the estimation for completeness but resulted in zero coefficients due to its high correlation with temperature ($\rho > 0.94$), consistent with the variable selection in the MBB-RF framework.

Overall, both the sparse VAR and MBB-RF models support a common narrative: drug demand is primarily shaped by its own past values and lagged temperature effects, with secondary contributions from other environmental factors. The convergence of results across linear (sparse VAR) and nonlinear (MBB-RF) models strengthens confidence in the robustness and generalizability of the identified predictors.

\begin{table}[h]
\centering
\caption{Non-zero coefficients for the drug demand equation in sparse VAR of order 
 4.\label{tab:lasso}}
\begin{tabular}{@{}lcccc@{}}
\toprule
\textbf{Variable} & \textbf{Lag 1} & \textbf{Lag 2} & \textbf{Lag 3} & \textbf{Lag 4} \\
\midrule
Temperature       & -0.1841 & 0       & -0.1167 & 0       \\
FWI               & -0.0251 & 0       & 0       & 0       \\
WindSpeed         & 0       & 0       & 0.0174  & -0.0018 \\
ExtremeRainfalls  & 0       & 0       & 0       & 0       \\
WetDays           & 0       & 0       & 0       & 0       \\
TemperatureSD     & 0       & 0       & 0       & -0.0201 \\
SpecificHumidity  & 0       & 0       & 0       & 0       \\
DrugDemand     & 0.2575  & 0.0139  & 0.1282  & 0.1433  \\
CloudCover        & 0       & 0.0257  & 0       & 0       \\
Precipitation     & 0       & 0.0024  & 0       & 0       \\
\bottomrule
\end{tabular}
\end{table}

\subsection{Forecasting models}
The forecasts produced by the univariate Prophet model serve as a benchmark for evaluating predictive performance, while the fitted values—representing smoothed versions of the endogenous variables—are included as additional covariates in all subsequent forecasting models. This strategy enables the models to leverage Prophet capacity to flexibly extract trend and seasonal components without modifying their underlying parametric structure. A rolling-window forecasting experiment further supported the inclusion of fitted series (see Supplementary material, Section~S6), confirming their contribution to improved forecast accuracy.

It is noteworthy that the Prophet model did not incorporate holiday effects as external regressors. An in-depth analysis revealed no statistically significant difference in drug demand between holiday and non-holiday weeks (see Supplementary material~S4), thus justifying their exclusion.

\subsubsection*{VARX MODEL}
In the multivariate VARX(4) model, where the lag order was selected based on the BIC and the estimation was carried out using a residual bootstrap procedure to ensure robust inference, the endogenous variables comprise temperature and respiratory drug demand. This specification is supported by the combined evidence from Granger causality spectra, variable importance analysis, and the sparse VAR model, all of which consistently identified temperature as the most influential climate-related predictor, alongside the strong autoregressive dynamics of the target variable.

A set of single-harmonic Fourier terms was included as exogenous regressors to account for the dominant seasonal component observed in the weekly data. This allowed the model to capture annual periodicity explicitly without requiring a higher lag order, thereby reducing model complexity while preserving explanatory power. Additionally, a binary dummy variable was introduced to account for the systematic drop in pharmaceutical sales during the week of August 15th—a national holiday period characterized by reduced healthcare activity, high temperatures, and consistently the lowest demand observed throughout the year.

Fitted values from univariate Prophet models for both endogenous variables were also included as exogenous inputs. While the main goal is to forecast drug demand, incorporating the smoothed Prophet series for temperature helps improve its forecast within the VARX structure. This enhanced representation of temperature propagates through the system via the multivariate dynamics of the model, ultimately yielding more accurate predictions of drug demand. The resulting hybrid strategy combines Prophet flexibility in capturing trend and seasonal components with the VARX ability to model lagged interdependencies, offering a robust and comprehensive framework for forecasting drug demand under climatic variability.

The results in Table~\ref{tab:varx_estimates} indicate that higher temperatures in the four weeks preceding each observation are generally linked to a decrease in drug demand, although only the third lag demonstrates a statistically significant effect. This finding aligns with the established seasonality of respiratory illnesses, which typically decline during warmer periods. Warmer weather is associated with improved air quality, increased outdoor activity, and lower transmission rates of respiratory pathogens. Conversely, colder temperatures can exacerbate respiratory conditions by increasing the spread of viral infections, reducing mucociliary clearance, and triggering bronchoconstriction, particularly in vulnerable populations. These mechanisms help explain the heightened drug demand during colder months and support the observed negative association between temperature and respiratory drug consumption. However, using weekly data may obscure the short-term impacts of extreme temperature events, such as heat waves, which can have acute and nonlinear effects on vulnerable populations. More detailed data (e.g., daily resolution) are therefore desirable to fully capture these dynamics and disentangle the effects of gradual seasonal changes from those of short-lived but intense climatic episodes, which are expected to become more frequent and severe in climate change.

According to the chosen estimation procedure, key assumptions regarding the residuals of the VARX model were assessed using empirical p-value approaches based on bootstrapped statistics, ensuring robustness against finite-sample distortions and model misspecification. Specifically, residual diagnostics included a Portmanteau test for serial correlation and an ARCH-LM test for conditional heteroskedasticity. Both tests were applied using 12 and 52 lags, corresponding approximately to a quarterly and an annual window, respectively, to capture potential short- to long-term volatility clustering in the weekly data.

Results were consistent across both lag specifications, showing no significant evidence of either serial correlation or heteroskedasticity, supporting the validity of the residual bootstrap inference framework and the overall reliability of the estimated VARX model.

Additionally, Granger causality tests in the time domain were conducted within the complete VARX system to evaluate the marginal contribution of each climatic predictor to drug demand over time. Although the model structure differs from the spectral analysis due to the inclusion of exogenous regressors, the results provide strong evidence that temperature Granger-causes drug demand, with an empirical p-value of 0.0001 based on bootstrap inference. The p-value was computed using a conservative correction \citep{phipson2010permutation}, which adds one to both the numerator and denominator of the empirical ratio, thereby avoiding the occurrence of zero p-values and yielding a more robust estimate in finite samples. While this confirms a highly significant predictive relationship, it does not provide any information on the frequency-specific structure of this link, which was previously revealed by the frequency-domain Granger causality spectra.

\begin{figure}[t!]
\centerline{\includegraphics[scale = 0.55]{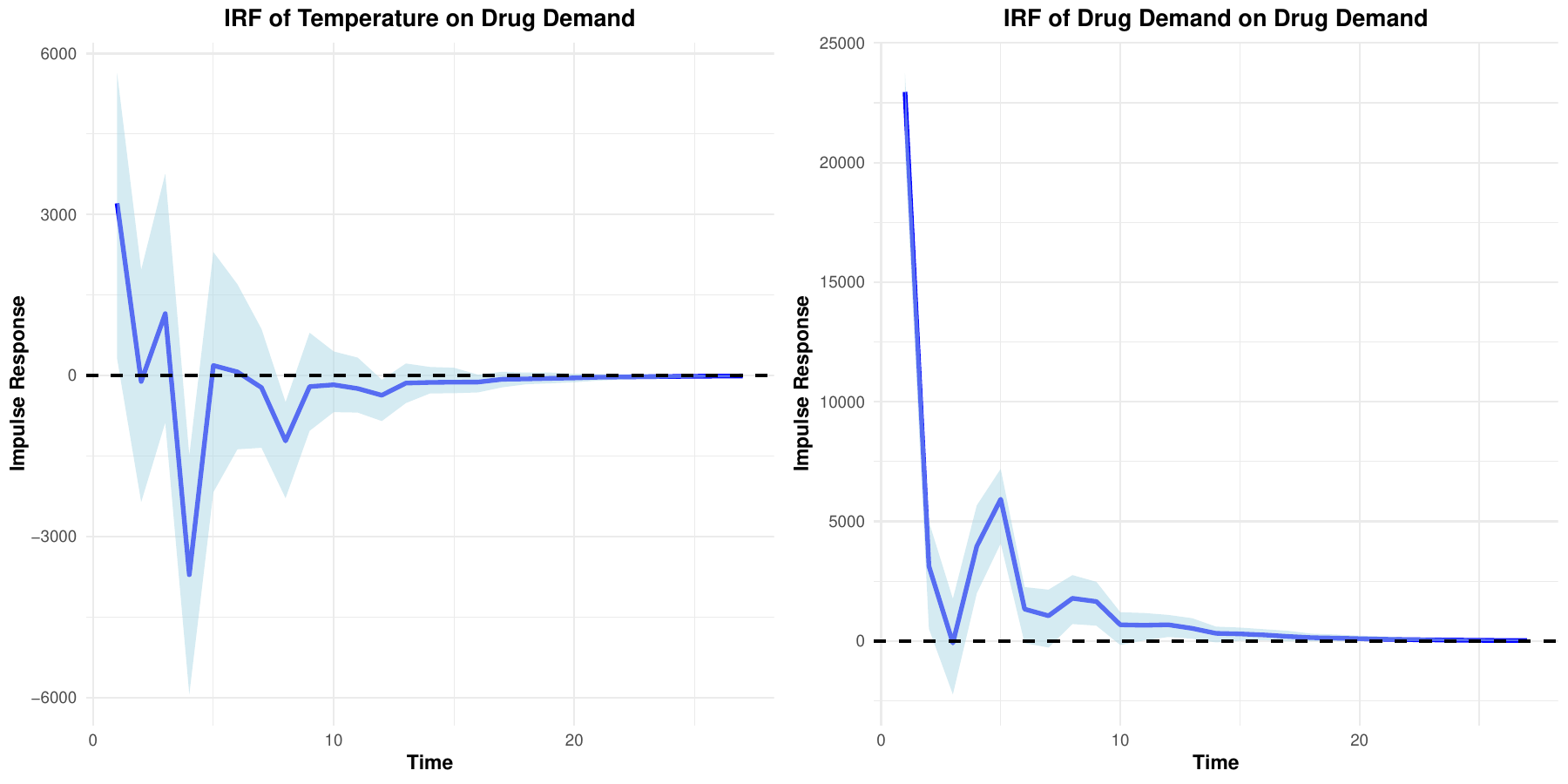}}
\caption{\textit{Impulse Response Functions (IRFs) estimated from a VARX(4) model with temperature and respiratory drug demand as endogenous variables. Exogenous regressors include a single-harmonic Fourier term, Prophet-fitted values of the endogenous variables, and a dummy variable for the national holiday on August 15th. Shaded areas represent 95\% confidence intervals obtained via the residual bootstrap method. The forecast horizon extends to 26 weeks (one semester). After 13 weeks (approximately one quarter), the responses flatten and become negligible.}} \label{fig:irf}
\end{figure}

The Impulse Response Functions (IRFs), displayed in Figure~\ref{fig:irf}, offer valuable insights into the dynamic relationship between temperature, past drug demand, and current prescription patterns. These IRFs are derived from the estimated VARX(4) model described in Table~\ref{tab:varx_estimates} to trace the effects of one-standard-deviation shocks on respiratory drug demand over time in each endogenous variable. 

The left panel in Figure~\ref{fig:irf} displays the response of drug demand to a one-standard-deviation shock in temperature. Although the confidence bands encompass zero across much of the horizon—likely due to the inclusion of strong seasonal controls such as Fourier terms and the mid-August dummy—statistically significant responses emerge in the early weeks. Specifically, the effect initially turns negative, suggesting that a sudden temperature increase may temporarily reduce demand, potentially due to behavioral adjustments (e.g., reduced indoor crowding or lower viral transmission in ventilated environments). This is followed, however, by a delayed increase in drug demand after the fourth week. 

The right panel in Figure~\ref{fig:irf} presents the IRF of drug demand to its own shock. The response confirms the strong autoregressive nature of the series, with an immediate and persistent effect that gradually diminishes over the 20-week horizon. This pattern is consistent with previous evidence from the MBB-RF importance analysis and aligns with the structure of the VARX model, both of which highlight the predictive role of past demand levels. 

The IRF analysis thus adds temporal nuance to the interpretation of climate-health relationships, capturing both immediate and lagged effects within a multivariate framework. These results are further supported by the Forecast Error Variance Decomposition (FEVD) reported in Table~S6 in the Supplementary material, quantifying the relative contribution of temperature shocks to the forecast error variance of drug demand. Although the majority of forecast error variance is, as expected, attributed to the autoregressive dynamics of drug demand itself (approximately 94\%), temperature consistently explains around 6\% of the variance from week 4 onward. The decomposition stabilizes by week 12 and remains remarkably constant over longer horizons, with 95\% confidence intervals for the temperature contribution ranging approximately from 2.2\% to 10.4\%.

\begin{table}[htbp]
\footnotesize
\centering
\caption{Bootstrap estimates and confidence intervals for the VARX model coefficients. Asterisks (*) indicate significance at the 5\% level based on 1,000 bootstrap replications.}
\label{tab:varx_estimates}
\begin{tabular}{llrrrc}
\toprule
\textbf{Equation} & \textbf{Coefficient} & \textbf{Estimate} & \textbf{Lower Bound} & \textbf{Upper Bound} & \\
\midrule
\textbf{Drug Demand}
& Constant & -67722.6388 & -153020 & 24610 & \\
& Temperature (t-1) & -342.0322 & -1939.9 & 1134.7 & \\
& Drug Demand (t-1) & 0.1363 & 0.0135 & 0.2096 & * \\
& Temperature (t-2) & 887.7639 & -803.26 & 2492.4 & \\
& Drug Demand (t-2) & -0.0206 & -0.1210 & 0.0650 & \\
& Temperature (t-3) & -3013.8000 & -4650.2 & -1468.7 & * \\
& Drug Demand (t-3) & 0.1719 & 0.0561 & 0.2556 & * \\
& Temperature (t-4) & 920.8347 & -653.29 & 2312.7 & \\
& Drug Demand (t-4) & 0.2236 & 0.1199 & 0.3093 & * \\
& Prophet fitted (Drug Demand) & 0.8742 & 0.6905 & 1.1375 & * \\
& August 15th & -36094 & -55318 & -18324 & * \\
& Prophet fitted (Temp) & 1106.3878 & -2592.3 & 5188.7 & \\
& $\sin(2\pi t / 52)$ & -20919.3932 & -40735 & -653.71 & * \\
& $\cos(2\pi t / 52)$ & -8007.0490 & -43405 & 26181 & \\
\midrule
\textbf{Temperature}
& Constant & -4.3895 & -9.9889 & 3.1394 & \\
& Temperature (t-1) & 0.3308 & 0.1875 & 0.4078 & * \\
& Drug Demand (t-1) & 4.87e-6 & -1.67e-6 & 1.24e-5 & \\
& Temperature (t-2) & 0.0361 & -0.1005 & 0.1302 & \\
& Drug Demand (t-2) & -1.17e-6 & -7.74e-6 & 5.92e-6 & \\
& Temperature (t-3) & -0.0186 & -0.1283 & 0.0852 & \\
& Drug Demand (t-3) & -4.29e-6 & -1.08e-5 & 2.09e-6 & \\
& Temperature (t-4) & 0.0978 & -0.0266 & 0.1755 & \\
& Drug Demand (t-4) & 4.12e-6 & -3.09e-6 & 1.05e-5 & \\
& Prophet fitted (Drug Demand) & 5.06e-7 & -1.77e-5 & 1.61e-5 & \\
& August 15th & -0.1689 & -1.6623 & 1.2397 & \\
& Prophet fitted (Temp) & 0.7814 & 0.5431 & 1.1008 & * \\
& $\sin(2\pi t / 52)$ & 1.5482 & -0.0743 & 2.5364 & \\
& $\cos(2\pi t / 52)$ & 1.4369 & -1.1401 & 3.5149 & \\
\bottomrule
\end{tabular}
\end{table}

\subsubsection*{Random Forest}
The same set of predictor variables was used across all models to ensure a fair comparison between the forecasting performance of the parametric VARX model and the nonparametric machine learning alternative represented by MBB-RF. Consistently with the feature importance analysis, the predictive MBB-RF was trained using 1,000 trees. Each tree was built on a resampled time series composed of overlapping blocks of 52 consecutive weeks, thereby preserving the seasonal structure inherent in the weekly data using a minimum node size of five.

The resulting model achieved an out-of-bag (OOB) RMSE of 24,460.15, along with a RSR of 0.6033 and an $R^2$ of 0.6360. The OOB metrics slightly outperform those obtained under a 52-week training–test split. This difference can be attributed to the nature of the MBB resampling scheme, which does not fully reproduce the temporal independence typically ensured by a strict separation between training and testing sets despite its statistical robustness. Specifically, the sampled blocks used to train each tree remain close in time to the OOB observations, often sharing similar seasonal patterns. As a result, the forecasting task becomes relatively easier than in a forward-chaining evaluation, where the model must extrapolate to unseen future periods that different environmental or behavioral dynamics may influence. In this context, test-set and rolling-window evaluation procedures offer a more conservative and realistic benchmark for assessing generalization performance under temporal dependence.

\begin{figure}[t!]
\centering
\includegraphics[scale=0.35]{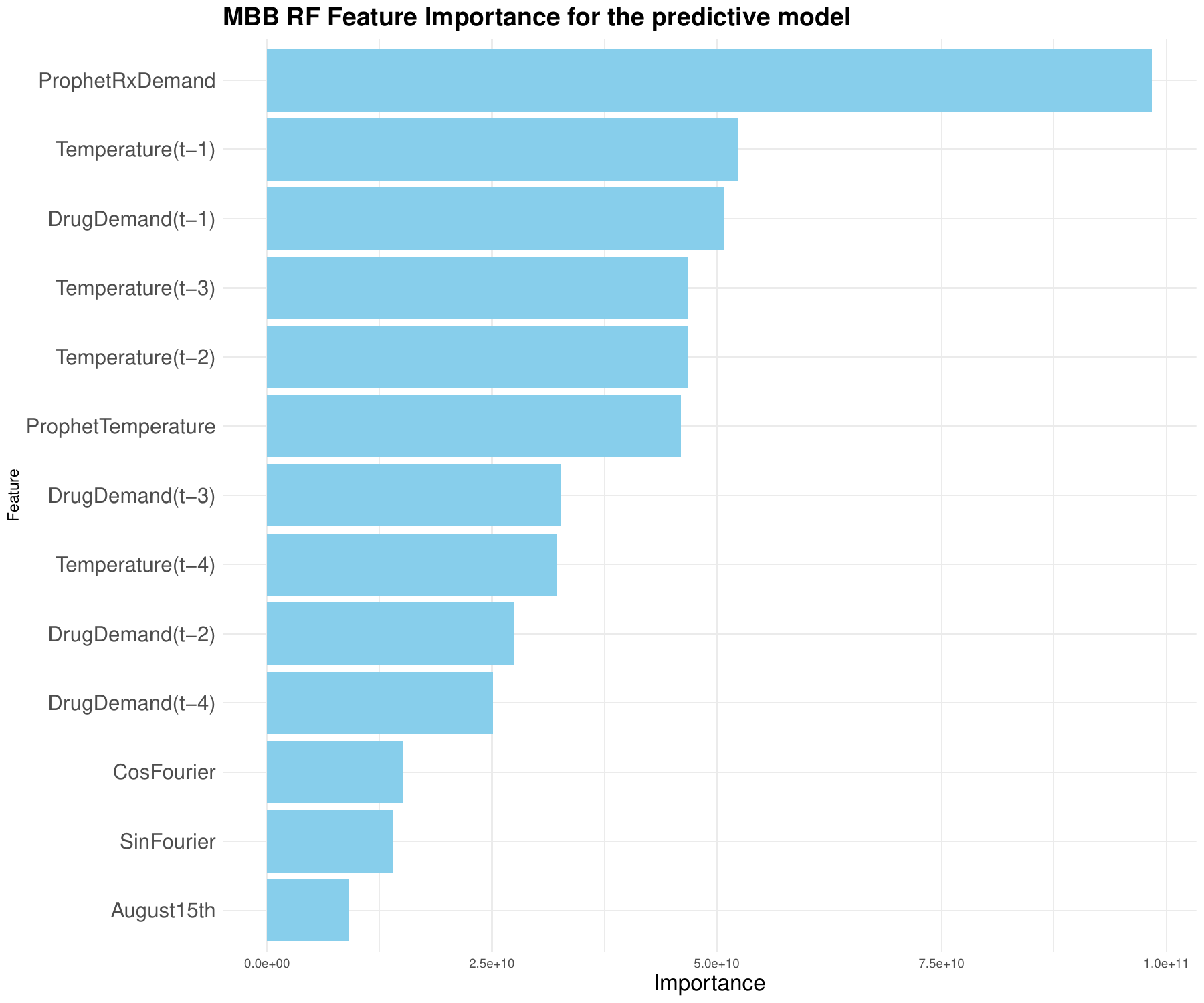}
\caption{\textit{Feature importance derived from the predictive Random Forest model trained under a Moving Block Bootstrap scheme (MBB-RF). Importance is measured by total impurity reduction across all trees. Variables are ranked in decreasing order of their contribution to forecasting weekly respiratory drug demand.}}
\label{fig:rf_imp_def}
\end{figure}

According to the feature importance results shown in Figure~\ref{fig:rf_imp_def}, the fitted values from Prophet rank at the top, suggesting that the structural components extracted by this model provide substantial predictive value when used as covariates.

These are followed by lagged values of drug demand and temperature, reinforcing the findings from Granger causality analysis and autoregressive modeling. Notably, the first lag of temperature and the Prophet-fitted values outperform even the first lag of the target variable, despite the known strong autoregressive behavior of the series. This may reflect the ability of these inputs to capture broader, smoothed temporal signals that complement the more localized yet potentially noisier autoregressive dynamics.

Seasonal regressors (namely, the Fourier harmonics and the August 15th dummy) appear lower in the ranking but still retain non-negligible importance, confirming the role of annual cyclicality in shaping weekly drug demand.

\subsubsection*{LSTM}
The LSTM architecture was implemented on an input consisting of sequences of 52 time steps, each including drug demand and temperature. An extensive grid search over key architectural and training hyperparameters was performed to identify the optimal neural network configuration. Specifically, multiple combinations of the number of LSTM units (40, 50, 80, 100), batch sizes (16, 32), the inclusion of attention mechanisms, and early stopping strategies were considered.Each configuration was trained using the Adam optimizer for up to 100 epochs, with early stopping based on the validation loss. A validation set comprising 10\% of the training data was used to monitor performance, which was evaluated based on the minimum validation Mean Squared Error (MSE) achieved during training.

The best-performing architecture consisted of 80 units, a batch size 32, no attention mechanism, and early stopping enabled. A 20\% dropout rate was reactivated during inference using Monte Carlo Dropout~\citep{srivastava2014dropout} to evaluate the stability and reliability of the LSTM forecasts (see Supplementary material, Section~S7).

\begin{figure}[t!]
\centering
\includegraphics[scale = 0.6]{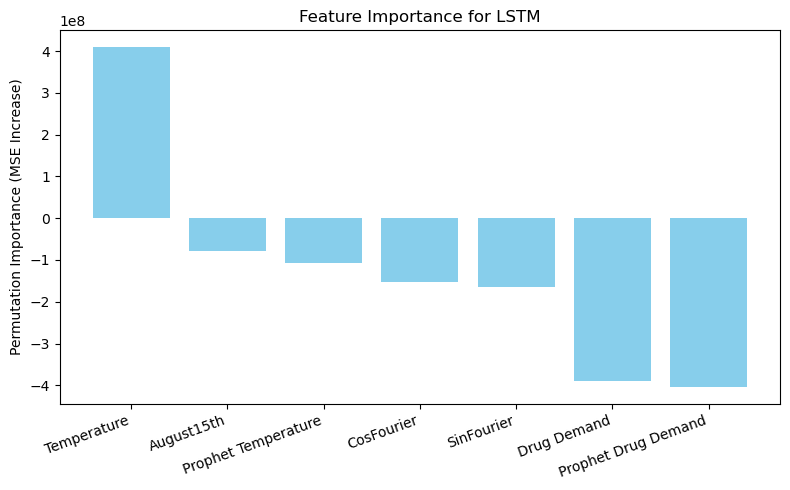}
\caption*{(a) Permutation importance}

\vspace{0.5cm}

\includegraphics[scale = 0.6]{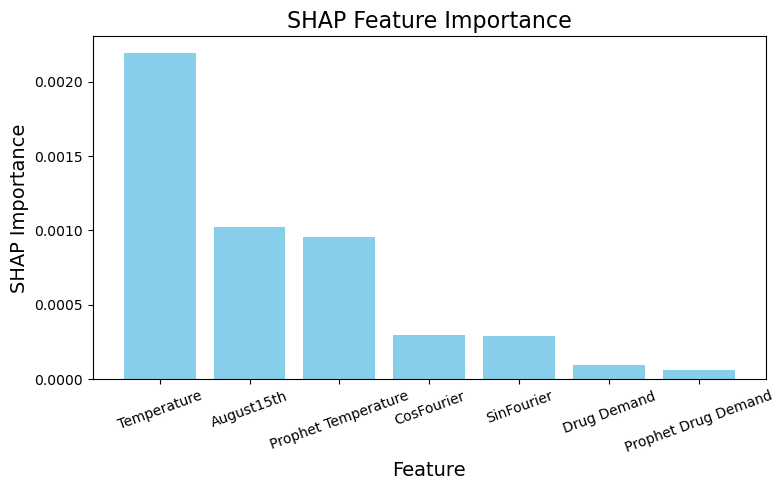}
\caption*{(b) SHAP values}

\caption{\textit{Model-agnostic interpretation of LSTM predictions. Panel (a) shows the permutation importance scores, measured by the increase in MSE when each feature is randomly shuffled. Panel (b) displays the average SHAP values, quantifying the marginal contribution of each predictor to the model output across the test set.}}
\label{fig:lstm_interpret}
\end{figure}

The permutation importance analysis (Figure~\ref{fig:lstm_interpret}a) and SHAP values (Figure~\ref{fig:lstm_interpret}b) offer complementary insights into the relative influence of each predictor in the LSTM model.

Across both interpretability frameworks, temperature emerges as the most important variable. Its high permutation score indicates that randomly shuffling this feature substantially degrades the model’s accuracy, confirming its key role in shaping respiratory drug demand. The SHAP values corroborate this finding, showing that temperature contributes the largest marginal effect to model predictions across the test set. 

The August 15th dummy and Prophet-fitted temperature are also relevant, particularly in the SHAP analysis. Their importance reflects the model sensitivity to known seasonal patterns and holiday-related drops in pharmaceutical activity. The Fourier terms exhibit moderate importance, suggesting that while they explicitly encode seasonality, part of this information may already be captured by the LSTM internal state, which is inherently suited to learning recurring temporal patterns from the input sequence.

In contrast, the permutation analysis reveals that Drug Demand and Prophet-fitted Drug Demand receive negative importance scores. This suggests that including these variables may introduce noise or redundancy, potentially degrading performance when not properly regularized. One explanation is that the autoregressive signal in drug demand may already be sufficiently captured by the model temporal dynamics and temperature-driven effects, making the explicit inclusion of these features unnecessary or even detrimental. The SHAP values for these same predictors are relatively small, indicating a limited contribution to individual predictions and further supporting their marginal role in the model.

Overall, the results suggest that environmental and seasonal exogenous features are more pivotal than the autoregressive component in shaping LSTM forecasts. This finding is particularly relevant in our interest in climate change, as it underscores the dominant influence of climatic conditions on drug demand. Nevertheless, to ensure a fair and consistent comparison across models, the same feature set was retained throughout all predictive experiments, thereby isolating model performance differences from input information differences.

\subsubsection*{Discussion}

\begin{table*}[h!]
\centering
\normalsize 
\caption{Forecasting performance of the four models—Prophet, VARX, MBB-RF, and LSTM—on one-year-ahead test data. Each model is estimated or trained on a training set of 338 weekly observations and evaluated on a 52-week horizon. All models use the same predictors, including autoregressive lags, seasonal Fourier terms, the August 15th dummy, and Prophet-fitted temperature and drug demand values as exogenous regressors.}
\label{tab:metrics}
\begin{tabular*}{\textwidth}{@{\extracolsep\fill}lccccc@{\extracolsep\fill}}
\toprule
\textbf{Model} & \textbf{MAPE} & \textbf{RMSE} & \textbf{RSR} & \textbf{$\text R^2$} & \textbf{MASE} \\
\midrule
Prophet   & 12.0000&  36924.92&  0.7395& 0.4531 &  1.1207  \\
VARX & 10.8102 & 33243.92 & 0.6658 & 0.5567 & 1.0081 \\
RF & 10.5861 & 33361.86 & 0.6681 & 0.5536 & 0.9892 \\
LSTM & 11.3413 & 30886.91 & 0.6246 & 0.6099 & 0.8897 \\
\bottomrule
\end{tabular*}
\end{table*}

We assess the one-year-ahead forecasting performance of four models using a harmonized set of predictors and a consistent rolling evaluation framework (Table~\ref{tab:metrics}). All models were trained on a 338-week window and evaluated over a 52-week forecast horizon. The input features include seasonal Fourier terms and Prophet-fitted values for temperature and drug demand. This standardized setup ensures that differences in performance can be attributed to the model architecture itself, rather than to variation in input data or training conditions.

Regarding relative accuracy, the MBB-RF model slightly outperforms all others, achieving the lowest MAPE (10.59\%) and a MASE below 1 (0.9892). These values indicate that MBB-RF effectively captures average demand patterns and consistently improves upon a seasonal naïve benchmark. Its $R^2$ value (0.5536) confirms a solid proportion of variance explained, supporting its ability to generalize across diverse temporal contexts.

The VARX model follows closely, with only marginally higher MAPE and MASE values. However, it achieves the lowest RMSE (33,243.92) and RSR (0.6658), highlighting its effectiveness in reducing large forecast deviations—an important property when the cost of extreme prediction errors is high. Its $R^2$ of 0.5567 is the second-highest, reaffirming that VARX balances accuracy and model stability. Additionally, its structural interpretability offers analytical advantages, as techniques such as IRFs and FEVD allow for the examination of lagged dynamics and propagation mechanisms across variables.

The LSTM model delivers the best overall performance on several metrics: RMSE (30,886.91), MASE (0.8832), and the highest $R^2$ (0.6099). These results underscore its capacity to learn long-range dependencies and model intricate temporal fluctuations. However, its MAPE (11.34\%) is slightly higher than those of RF and VARX, suggesting a trade-off: while LSTM excels in minimizing squared error and explaining variance, it may be more sensitive to shorter training sequences or overfitting. Furthermore, its black-box nature limits interpretability, requiring post-hoc methods like SHAP or LIME to extract meaningful insights.

Despite its proficiency in modeling trends and seasonality, the Prophet model ranks lowest across all metrics. With a MAPE of 12.00\%, MASE of 1.1207, and a relatively low $R^2$ (0.4531), Prophet struggles to match the performance of more flexible or feature-rich approaches. Its lack of climatic covariates hampers responsiveness to environmental drivers, particularly critical in this application where temperature, the leading climatic driver, substantially improves performance, notably in RMSE, MASE, and $R^2$.

Importantly, only MBB-RF and LSTM yield MASE values below 1, suggesting that these models consistently outperform a seasonal naïve benchmark. This reinforces their practical reliability, especially when evaluated using normalized accuracy metrics. In contrast, the slightly higher MASE of the VARX model—despite its strong RMSE and $R^2$ scores—suggests that its forecasts do not consistently surpass the accuracy of a simple seasonal repetition. This outcome may be attributed to a strongly seasonal demand pattern and limited abrupt climatic shifts within the observed period. In such cases, a naïve seasonal forecast (e.g., using last year’s values) may perform reasonably well, setting a high bar for MASE-based comparisons. Nonetheless, VARX remains valuable for its ability to capture and interpret lagged dependencies and for offering transparent insights into long-run interactions, particularly when environmental variability is subtle and structural understanding is prioritized over marginal gains in relative accuracy.

These findings underscore the importance of integrating climate-sensitive covariates into forecasting models. Approaches that combine temporal dependencies with exogenous regressors, particularly those capable of modeling nonlinear interactions or complex sequences, consistently outperform simpler autoregressive alternatives. However, the robust performance of VARX demonstrates that well-specified parametric structures can remain competitive, especially when model transparency and causal interpretability are essential. 

The final choice among models reflects trade-offs among predictive accuracy, interpretability, and model complexity. This balance depends heavily on the domain context of forecast-driven decisions and policy relevance. When the primary objective is predictive accuracy, particularly in minimizing large forecast deviations, the LSTM model stands out as the most effective choice. For applications where average predictive performance is prioritized, VARX and MBB-RF emerge as strong alternatives, achieving lower MAPE values. Among these, VARX is particularly well suited when interpretability and the analysis of temporal dynamics are essential, due to its parametric structure and the availability of tools such as impulse response functions and forecast error variance decomposition. In contrast, MBB-RF may be preferable in contexts where nonlinear relationships and complex feature interactions are expected, offering greater modeling flexibility. The final model selection should reflect the trade-offs between forecast precision, transparency, and the need for structural insight, depending on the specific goals of the study.

In summary, when accuracy under large deviations is paramount, LSTM is preferable; when average precision and transparency are valued, VARX is ideal; and in contexts where complex interactions may play a role, MBB-RF offers a flexible and robust alternative.

\section{Concluding remarks}
This study implements and compares a range of forecasting models to predict the weekly demand for prescription pharmaceuticals in Greece, with the broader goal of quantifying how climate and environmental conditions influence medical needs in the population. As climate change intensifies—bringing more frequent and extreme weather events—anticipating shifts in drug demand becomes increasingly important for public health planning and supply chain resilience.

We first apply a joint analysis of unconditional and conditional Granger causality spectra to understand which environmental factors matter most. This frequency-domain approach reveals that temperature, specific humidity, wind speed, and the FWI exhibit the strongest and most consistent associations with respiratory pharmaceutical demand, particularly at low and medium frequencies. Crucially, these variables maintain statistical significance even when controlling for other climate-related predictors, emerging as prime candidates for inclusion in forecasting models.


However, the Granger causality framework inherently limits the analysis to pairwise or triplet relationships, making it difficult to assess the joint influence of multiple predictors. To overcome this limitation, we complement the Granger-based findings with a Random Forest model trained using a Moving Block Bootstrap scheme (MBB-RF), which enables simultaneous evaluation of all predictors while accounting for temporal dependence. The MBB-RF importance rankings corroborate the Granger results: temperature, specific humidity, and the Fire Weather Index (FWI) consistently emerge as the most influential predictors of pharmaceutical demand. 

The sparse VAR with Lasso penalization further supports the predominance of temperature, where the largest coefficients are associated with lags 1 and 3 of temperature, followed by lag 2 of cloud cover and lag 1 of FWI. This validates and reinforces the findings obtained through the other methodologies.

Nevertheless, the importance distribution exhibits a pronounced elbow, indicating a sharp drop in variable relevance, measured in terms of impurity reduction, after accounting for temperature and the autoregressive terms of drug demand. Based on this observation, and to reduce redundancy, only temperature and the lagged values of the target variable were retained in the final predictor set. This parsimonious selection balances model simplicity and forecasting performance, capturing both the intrinsic temporal dynamics of respiratory pharmaceutical use and the most impactful climatic signal.

Performance metrics for the implemented forecasting models highlight that incorporating temperature as a climatic regressor consistently enhances forecasting accuracy across all model classes. Its seasonal behavior reflects not only climatic rhythms but also behavioral and epidemiological dynamics related to the spread of respiratory illnesses, making it an effective predictor of demand fluctuations.

Among the evaluated models, the VARX structure that included Prophet-fitted exogenous inputs initially offered a good balance between interpretability and predictive accuracy based on squared error metrics. However, the MBB-RF outperformed it in terms of relative error measures such as MAPE and MASE, while also enabling variable importance analysis that supported the findings from the Granger causality framework. The LSTM model, although slightly less accurate in relative terms, achieved the best performance in RMSE and RSR, highlighting its strength in capturing nonlinear dynamics and high-frequency temporal patterns when sufficient structure is available in the data.


Taken together, the evidence points to a clear and concerning conclusion: climatic conditions, especially temperature, have a causal impact on the demand for respiratory medication. As climate variability grows, this relationship is likely to intensify, posing new challenges for healthcare systems. Forecasting models must therefore be equipped to anticipate these shifts, not only to optimize supply chains, but also to inform broader health preparedness strategies in a warming world.

Another avenue for future development concerns the temporal scope and resolution of the data. In this study, we relied on seven and a half years of weekly data, which is suitable for identifying recurring seasonal patterns and medium-term variability. However, to more robustly assess the long-term effects of climate change, which unfold over decades, it would be valuable to extend the historical coverage of the dataset.

Moreover, adopting a finer temporal resolution, such as daily data, could improve the model’s ability to capture the effects of short-lived but intense events, including heat waves, extreme rainfall, and other forms of acute environmental stress that are becoming increasingly frequent due to climate change. This would enable a more detailed analysis of how climate extremes affect medical needs on shorter timescales, complementing the broader patterns captured by weekly aggregates.

Further research could also expand the analysis to include a spatial dimension, integrating geographic heterogeneity and localized climate exposure. These enhancements could further support climate-aware planning in public health logistics, helping ensure that medication supply chains remain responsive in the face of a rapidly changing environment.

\section*{Acknowledgments}
The authors sincerely thank Dr. Filippo Benetti and Dr. Rosa Falotico of Alira Health Srl for their support in the acquisition and interpretation of the pharmaceutical data. The dataset used in this study is confidential and cannot be publicly shared.

\subsection*{Author contributions}

Viviana Schisa and Matteo Farnè contributed to the conception and design of the work. Viviana Schisa contributed to the data management, statistical analysis and interpretation of results. Viviana Schisa and Matteo Farnè contributed to the methodology and investigation. Viviana Schisa drafted the work, and all the authors reviewed it for important intellectual content. All the authors approved the publication of the final version of the manuscript. All the authors agreed to be accountable for all aspects of the work to ensure that questions related to the accuracy or integrity of any part of the work are appropriately investigated and resolved.

\subsection*{Financial disclosure}

Viviana Schisa acknowledges financial support from Alira Health and the European Union – NextGenerationEU, under the Italian National Recovery and Resilience Plan (PNRR), Mission 4, Component 2, Investment 3.3, as part of the research fellowship Ex D.M. 352/2022.

\subsection*{Conflict of interest}

The authors declare no potential conflict of interests.

\section*{Supporting information}
Additional materials supporting the findings of this study are provided in the online Supplementary Information. These include detailed descriptive statistics and distributional diagnostics for all climatic and pharmaceutical variables used in the analysis, along with diagnostic plots such as density curves, boxplots, and autocorrelation functions. Univariate time series models are presented to characterize the seasonal and stochastic structure of respiratory drug demand, including the estimation of a SARIMA--eGARCH model with volatility and asymmetry diagnostics.

Further sections provide extensive validation of the multivariate VARX framework, including normality and cointegration tests, rolling window forecasting results, and forecast accuracy comparisons across models. The impact of including Prophet-fitted series as external regressors is evaluated, and additional analyses explore model robustness under temporal shifts.

Finally, a map of data collection sites across Greece is included to document the spatial coverage of the dataset. Locations correspond to the most populous city in each administrative region, ensuring alignment with pharmaceutical consumption patterns.

\bibliographystyle{apalike}
\nocite{*}
\bibliography{Climate_change_impact_drugs}

\clearpage
\renewcommand{\thesection}{S\arabic{section}}
\setcounter{section}{0}
\renewcommand{\thefigure}{S\arabic{figure}}
\renewcommand{\thetable}{S\arabic{table}}
\setcounter{figure}{0}
\setcounter{table}{0}

\section*{Supplementary Material}
This Supplementary Material provides additional figures, statistical tests, and robustness checks that complement the main analysis. The goal is to ensure transparency and reproducibility, and to offer further insight into the temporal, distributional, and structural characteristics of the data and models used.

\section{Geographical Distribution of the Study Regions}

\begin{figure}[h!]
    \centering
    \includegraphics[width=0.7\textwidth]{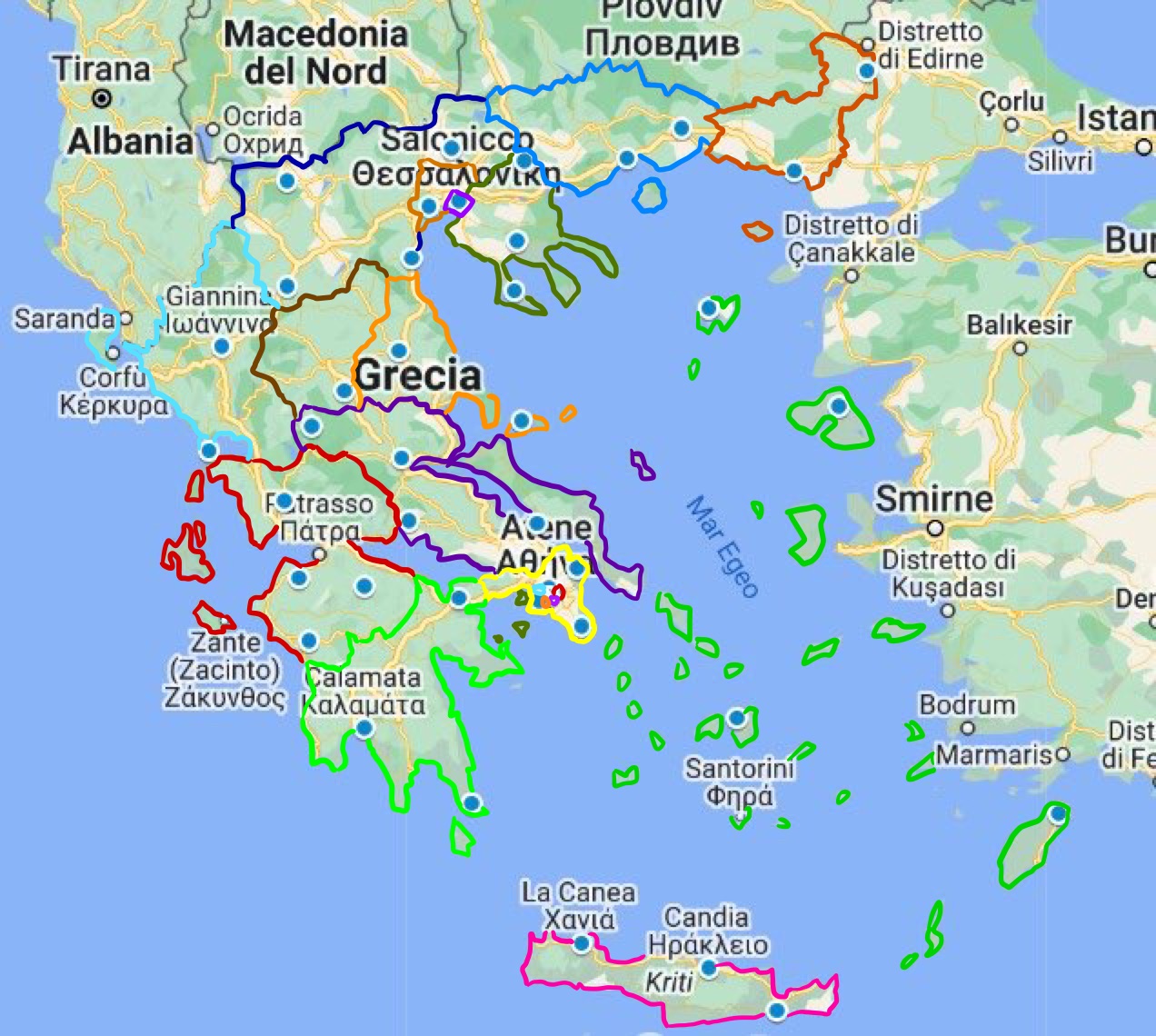}
    \caption{Map of the twenty pharmaceutical regions where Alira Health collects data on pharmaceutical consumption. For each region, the most populous city or urban location is marked with a blue dot and considered the most representative in terms of pharmaceutical consumption patterns.}
    \label{fig:map}
\end{figure}

Figure~\ref{fig:map} and Table~\ref{tab:regions} present the twenty pharmaceutical regions defined by Alira Health for collecting data on prescription drug consumption across Greece. For each region, we provide a representative city or urban location along with its coordinates. These reference points were used to geolocate environmental data in the analysis. We refer to \textit{cities or urban locations} because Athens is subdivided into four pharmaceutical regions, each represented by a different neighborhood. Although these areas fall within the same metropolitan context, they are treated as distinct regions in the dataset provided by Alira Health.

\begin{table}[h!]
\centering
\caption{Reference points for the twenty pharmaceutical regions.}
\label{tab:regions}
\begin{tabular}{|l|l|l|}
\hline
\textbf{Territory} & \textbf{City/Urban location} & \textbf{Coordinates} \\
\hline
Athens Central East & Ampelokipoi & 37.98N, 23.75E \\
Athens Central North & Patisia & 38.02N, 23.73E \\
Athens South & Neos Kosmos & 37.95N, 23.72E \\
Athens West & Petralona & 37.96N, 23.70E \\
Piraeus & Parnassidos & 37.93N, 23.63E \\
Rest of Attica & Maratona & 38.20N, 24.01E \\
Central Salonica & Salonica & 40.64N, 22.95E \\
Rest of Salonica East & Axios & 40.53N, 22.72E \\
Rest of Salonica West & Polygyros & 40.37N, 23.44E \\
Central Western Macedonia & Kozani & 40.29N, 21.79E \\
Eastern Macedonia & Drama & 41.15N, 24.15E \\
Thrace & Alexandroupoli & 40.86N, 25.87E \\
Central Greece & Lamia & 38.90N, 22.41E \\
Western Greece & Patrasso & 38.24N, 21.75E \\
Thessaly East & Larissa & 39.63N, 22.41E \\
Thessaly West & Karditsa & 39.36N, 21.92E \\
Eastern Peloponnese & Nafplio & 37.57N, 22.81E \\
Epirus & Ioannina & 39.66N, 20.88E \\
Crete & Irakleio & 35.32N, 25.15E \\
Aegean & Rhodes & 36.29N, 28.04E \\
\hline
\end{tabular}
\end{table}

\section{Descriptive analysis}
\begin{table}[h!]
\caption{Descriptive statistics of weekly variables aggregated at the national level. Variables are expressed in the following units: drug demand (number of prescription packages), temperature (°C), wind speed (m/s), cloud cover (proportion, 0–1), specific humidity (g/kg), precipitation (mm), Fire Weather Index (unitless), temperature standard deviation (°C), extreme rainfall (mm), and wet days (count of days with precipitation $>$ 1 mm). All variables are computed weekly and then spatially aggregated across regions using the mean, except for precipitation and extreme rainfall, which are aggregated using the sum due to their cumulative nature. Alira Health provides drug sales data, while climate variables are obtained from the ERA5 reanalysis dataset through the Copernicus Climate Data Store.}
\label{tab:desc_stats}
\centering
\begin{tabular}{lrrrrrr}
\toprule
\textbf{Variable} & \textbf{Min} & \textbf{1st Quartile} & \textbf{Median} & \textbf{Mean} & \textbf{3rd Quartile} & \textbf{Max} \\
\midrule
Drug Demand        & 79787   & 169350   & 196648   & 200144   & 228925   & 334525 \\
Temperature    & 1.68    & 9.80     & 15.11    & 15.63    & 22.25    & 29.05 \\
Wind Speed    & 1.46    & 2.07     & 2.38     & 2.42     & 2.76     & 4.07 \\
Cloud Cover   & 0.0133  & 0.2267   & 0.4201   & 0.3985   & 0.5429   & 0.8767 \\
Specific Humidity  & 2.8180 & 5.9170 & 7.4110 & 7.7280 & 9.8390 & 12.9630 \\
Precipitation  & 0.34    & 64.78    & 172.79   & 262.27   & 385.67   & 1510.49 \\
FWI                & 1.00    & 1.84     & 4.87     & 15.46    & 27.40    & 79.13 \\
Temperature SD & 0.405   & 0.997    & 1.342    & 1.476    & 1.787    & 4.812 \\
Extreme Rainfall  & 0.0  & 0.0      & 0.0      & 79.1     & 59.8     & 1422.4 \\
Wet Days   & 0.0     & 0.7      & 1.52     & 1.79     & 2.6      & 6.15 \\
\bottomrule
\end{tabular}
\end{table}

\begin{figure}[h!]
    \centering
    \includegraphics[scale = 0.5]{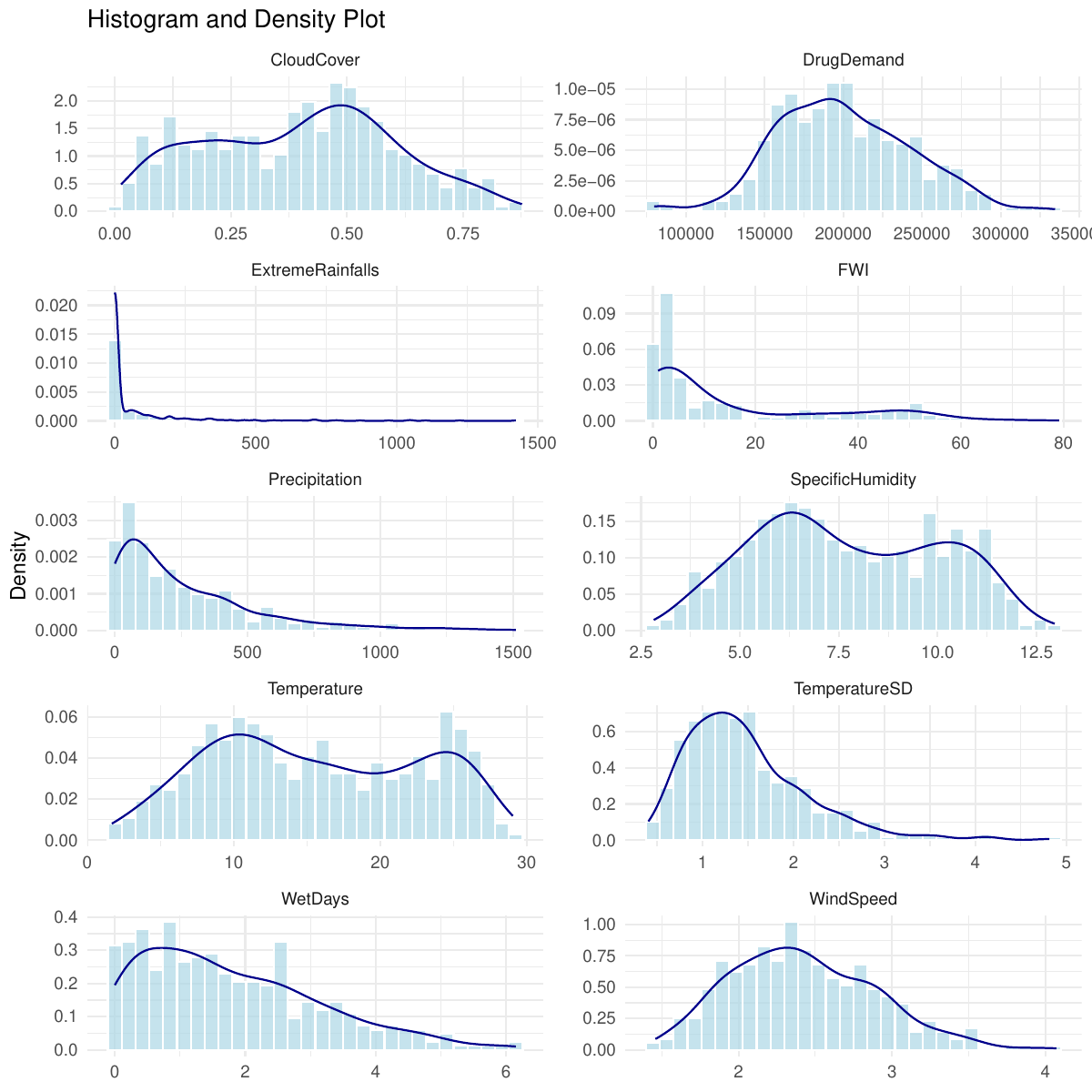}
    \caption{Density plots for the weekly variables, illustrating the distributional shape and skewness.}
    \label{fig:density}
\end{figure}

\begin{figure}[h!]
    \centering
    \includegraphics[scale = 0.45]{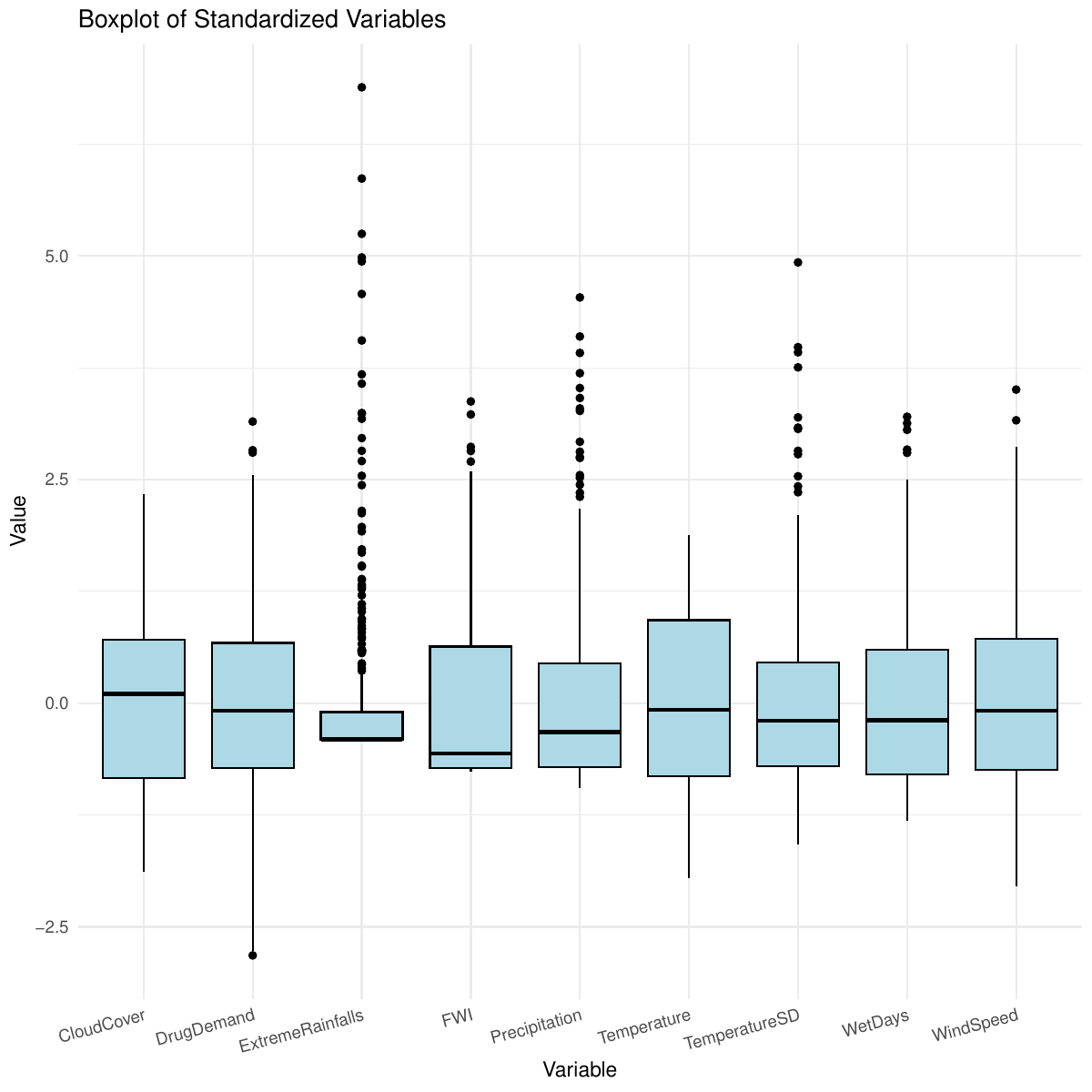}
    \caption{Boxplots for the standardized weekly variables, highlighting dispersion and presence of outliers.}
    \label{fig:boxplot}
\end{figure}

The descriptive statistics presented in Table~\ref{tab:desc_stats} provide an overview of the weekly variables included in the study. Figures~\ref{fig:density} and~\ref{fig:boxplot} complement this summary by visualizing the empirical distributions of each variable. The density plots highlight distinct distributional shapes, including multimodal or skewed behaviors, while the boxplots reveal the presence of extreme values. Notably, some variables (e.g., the Fire Weather Index and extreme rainfall) display long-tailed distributions, reflecting rare but intense events. While these events may influence pharmaceutical demand, their full impact might be underestimated at a weekly resolution. A finer temporal granularity, such as daily data, could allow for a more accurate capture of short-lived environmental shocks and their immediate consequences on drug consumption.

Shapiro–Wilk normality tests were conducted on each weekly time series to assess deviations from the Gaussian distribution. The results indicate significant departures from normality ($p < 0.001$) for all variables, except drug demand, which nonetheless shows a statistically significant result at the conventional 5\% level ($p = 0.0172$). These findings support using nonparametric or distributionally robust methods in subsequent analyses.

\begin{figure}[ht]
    \centering
    \includegraphics[width=\textwidth]{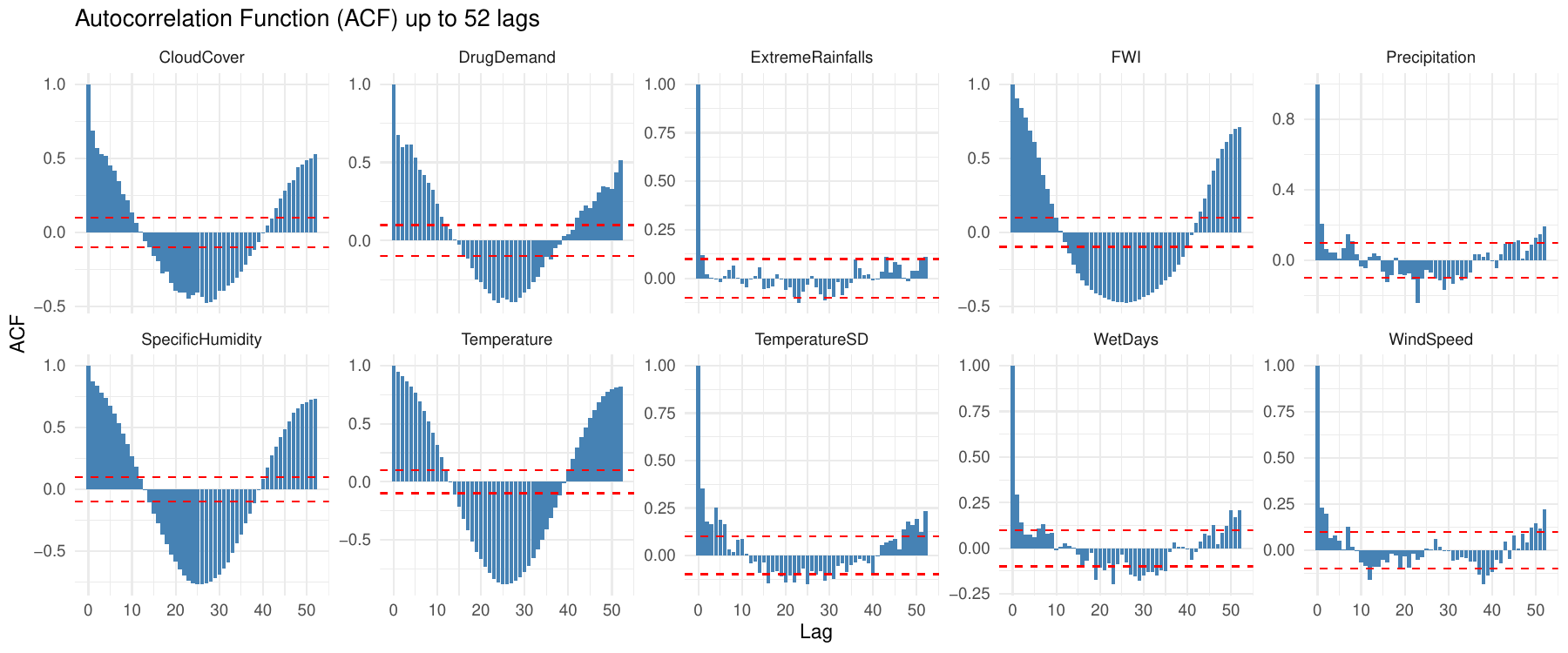}
    \caption{Autocorrelation Function (ACF) plots for each of the ten weekly time series up to 52 lags. The dashed red lines represent the $ \pm 1.96/\sqrt{T} $ confidence bounds, where $T$ is the number of observations. The presence of significant autocorrelation beyond lag zero in several series (notably in DrugDemand, Temperature, and FWI) supports the existence of temporal dependencies and justifies the application of time series models. These autocorrelations also shows cyclical patterns, motivating frequency-domain investigations.}
    \label{fig:acf_all}
\end{figure}

\section{Univariate data analysis}
\label{supplementary_data_analysis}
\begin{figure}[!h]
    \centering
    \includegraphics[scale = 0.55]{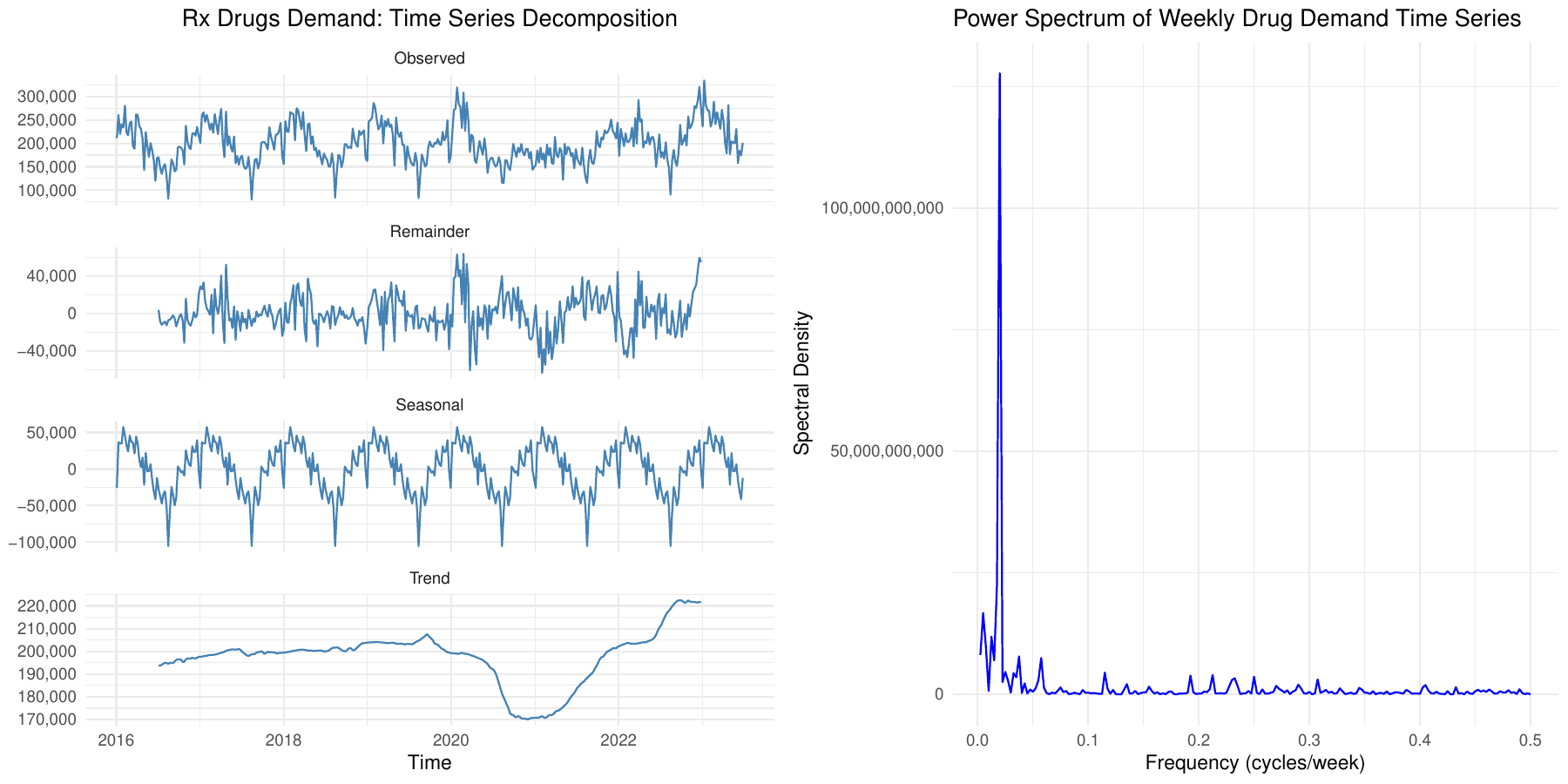}
    \caption{\textit{Time series decomposition and power spectrum of weekly respiratory drug demand (top) and average temperature (bottom). Left: The seasonal components clearly exhibit an annual cycle, reflecting both epidemiological and climatic rhythms. Right: The power spectra display marked low-frequency peaks near 0.019 cycles/week, consistent with a dominant yearly periodicity in both series.}}
    \label{fig:decomp_spec}
\end{figure}

Respiratory drug demand and average temperature share annual periodicity (see Figure~\ref{fig:decomp_spec}). The decomposition of pharmaceutical consumption highlights a strong seasonal signal, and its power spectrum displays a dominant low-frequency peak around 0.019 cycles/week, corresponding to a 52-week cycle. A similar, though more sharply defined, pattern emerges in the temperature series, where the seasonal dynamics follow the regular alternation of climatic conditions, and a single, narrow peak at the same frequency characterizes the spectrum.

In contrast, while still dominated by the annual signal, the spectrum of drug demand also shows minimal power at slightly higher frequencies. These negligible components may reflect minor irregularities that introduce modest deviations from the otherwise regular seasonal rhythm.

Beyond seasonality, the decomposed trends provide additional insight into the long-term behavior of the two series. For respiratory drug demand, the trend remains relatively stable during the pre-pandemic period, drops markedly with the onset of COVID-19, and stays low during the pandemic years—likely due to reduced transmission of respiratory infections following non-pharmaceutical interventions (e.g., lockdowns, mask mandates). Toward the end of the series, the trend resumes an upward trajectory, possibly reflecting a post-pandemic rebound in infections and the gradual return to routine healthcare practices.

In comparison, the temperature trend appears smoother and more stable over time. Although minor fluctuations are present, no abrupt changes are evident. This gradual evolution is consistent with the slow-moving nature of climatic change and highlights the greater inertia of environmental variables compared to the socio-behavioral sensitivity of pharmaceutical consumption.

The drug demand time series was first modeled univariately to gain insight into its temporal dynamics. The objective was to identify both deterministic seasonal components and stochastic features such as time-varying volatility. 

Let $y_t$ denote the weekly respiratory drug demand at time $t$, for $t = 1, \dots, T$, the conditional mean was modeled using a Seasonal ARIMA process, specifically $\text{SARIMA}(4,0,0)(1,0,0)_{52}$, selected through BIC minimization~\citep{dagum2001analisi}.
\begin{equation}
\left(1 - \phi_1 L - \phi_2 L^2 - \phi_3 L^3 - \phi_4 L^4\right)
\left(1 - \Phi_1 L^{52}\right) y_t = \varepsilon_t
\end{equation}
Here, $L$ denotes the lag operator, $\phi_i$ are the coefficients of the non-seasonal autoregressive terms, $\Phi_1$ is the seasonal autoregressive coefficient at lag 52, and $\varepsilon_t$ is a white-noise innovation. This specification effectively captures intra-annual regularities consistent with the seasonal pattern of respiratory illnesses.

\begin{figure}[t!]
    \centering
    \includegraphics[scale = 0.55]{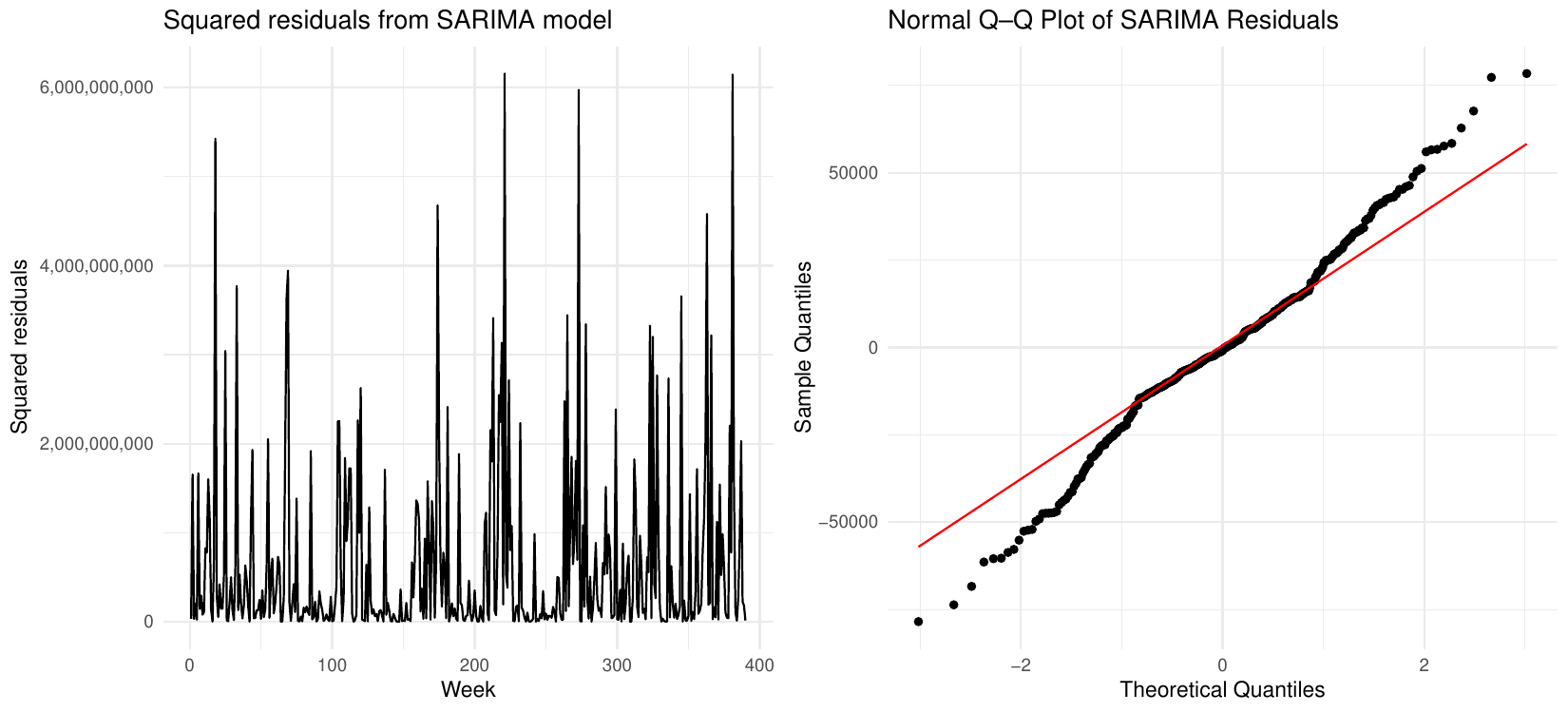}
    \caption{\textit{Residual diagnostics for the $\text{SARIMA}(4,0,0)(1,0,0)_{52}$ model for drug demand. Left panel: Time series of squared residuals, highlighting episodes of volatility clustering indicative of conditional heteroscedasticity. Right panel: Normal QQ plot of raw residuals, showing heavy tails and deviations from normality, especially in the upper and lower quantiles. These features support a volatility model with heavy-tailed and possibly skewed innovations.}}
    \label{fig:qqplot_sarima}
\end{figure}

The Ljung–Box test on the residuals yielded a p-value of 0.1239, indicating no significant remaining autocorrelation. However, signs of heteroscedasticity were evident: the ARCH-LM test was already significant at lag 1 ($p = 0.0427$), and the residuals deviated from normality, as indicated by the Shapiro–Wilk test ($p = 0.0011$). Figure~\ref{fig:qqplot_sarima} illustrates these features, revealing volatility clustering and non-Gaussian behavior with heavy tails and asymmetry. These empirical properties motivate the inclusion of a time-varying volatility component.

An Exponential GARCH model (eGARCH) to the residuals of the SARIMA model was fitted 
to account for volatility clustering and potential asymmetry in the conditional variance. The innovations were assumed to follow a Student-t distribution, allowing for fat tails. Specifically, an $\text{eGARCH}(1,1)$ model was estimated, following~\cite{francq2019garch} and~\cite{nelson1991conditional}:

\begin{equation}
    \log (\sigma_t^2) = \omega + \beta \log(\sigma_{t-1}^2) + \alpha \left| \frac{\varepsilon_{t-1}}{\sigma_{t-1}} \right| + \gamma \frac{\varepsilon_{t-1}}{\sigma_{t-1}}.
\end{equation}

The fitted model provides a satisfactory representation of the conditional variance dynamics. The volatility persistence is high, as indicated by $\hat{\beta} \approx 0.89$, suggesting that shocks have long-lasting effects. The ARCH parameter $\hat{\alpha} \approx 0.037$ is small and statistically insignificant, implying that immediate volatility responses to large shocks are limited. In contrast, the leverage term $\hat{\gamma} \approx 0.29$ is highly significant ($p < 0.001$), indicating asymmetry: volatility reacts more strongly to negative shocks than to positive ones.

\begin{table}[h!]
\centering
\caption{\textit{Estimated parameters of the $\text{eGARCH}(1,1)$ model with Student-t innovations fitted to the residuals of the $\text{SARIMA}(4,0,0)(1,0,0)_{52}$ model.}}
\label{tab:egarch_params}
\begin{tabular}{lcccc}
\toprule
\textbf{} & \textbf{Estimate} & \textbf{Std. Error} & \textbf{t value} & \textbf{Pr($>|t|$)} \\
\midrule
$\omega$   & 2.1983 & 0.1089 & 20.1859 & $<$0.0001 \\
$\alpha$ & 0.0371 & 0.0492 & 0.7540  & 0.4509 \\
$\beta$  & 0.8915 & 0.0055 & 161.4988 & $<$0.0001 \\
$\gamma$ & 0.2906 & 0.0751 & 3.8685  & $<$0.0001 \\
\bottomrule
\end{tabular}
\end{table}

This asymmetry may reflect behavioral dynamics in pharmaceutical consumption. Negative shocks—such as abrupt declines in demand—may signal disruptions in access, policy interventions, or shifts in health communication, all of which can introduce additional uncertainty. Conversely, demand increases often follow seasonal flu waves and are thus more predictable, contributing less to volatility. Importantly, the asymmetry does not imply that every negative shock increases variance, but rather that the system responds more sensitively to such deviations when volatility changes.

The COVID-19 pandemic illustrates this nuance (see Figure 1 in main article). Following the outbreak, demand for respiratory drugs dropped markedly and the series flattened, with a notable reduction in volatility. This likely reflects the impact of coordinated public health measures (e.g., lockdowns, mask mandates) that simultaneously reduced infections and stabilized demand patterns. While this shock decreased both the level and variability of demand, it also disrupted expected seasonal dynamics, supporting the model's ability to detect structural changes in volatility behavior.

The shape parameter is 6.24, indicating moderately heavy tails, supporting the choice of a Student-t distribution over a Gaussian alternative.

Model diagnostics confirm the adequacy of the specification. Weighted Ljung–Box tests on standardized and squared residuals yield high $p$-values ($> 0.5$), suggesting no remaining serial correlation or unmodeled conditional heteroscedasticity. Moreover, weighted ARCH-LM tests are not significant, reinforcing the conclusion that volatility clustering is adequately captured. Sign bias tests are also non-significant, indicating that no asymmetry remains unaccounted for in the residuals.

Finally, while the adjusted Pearson's goodness-of-fit test reports significant deviations for group sizes 20 and 50 ($p < 0.05$), this may reflect mild distributional misspecification—an expected behavior when modeling heavy-tailed series. The model delivers a robust and well-calibrated representation of the weekly respiratory drug demand volatility structure.

\begin{figure}[t!]
    \centering
    \includegraphics[scale = 0.55]{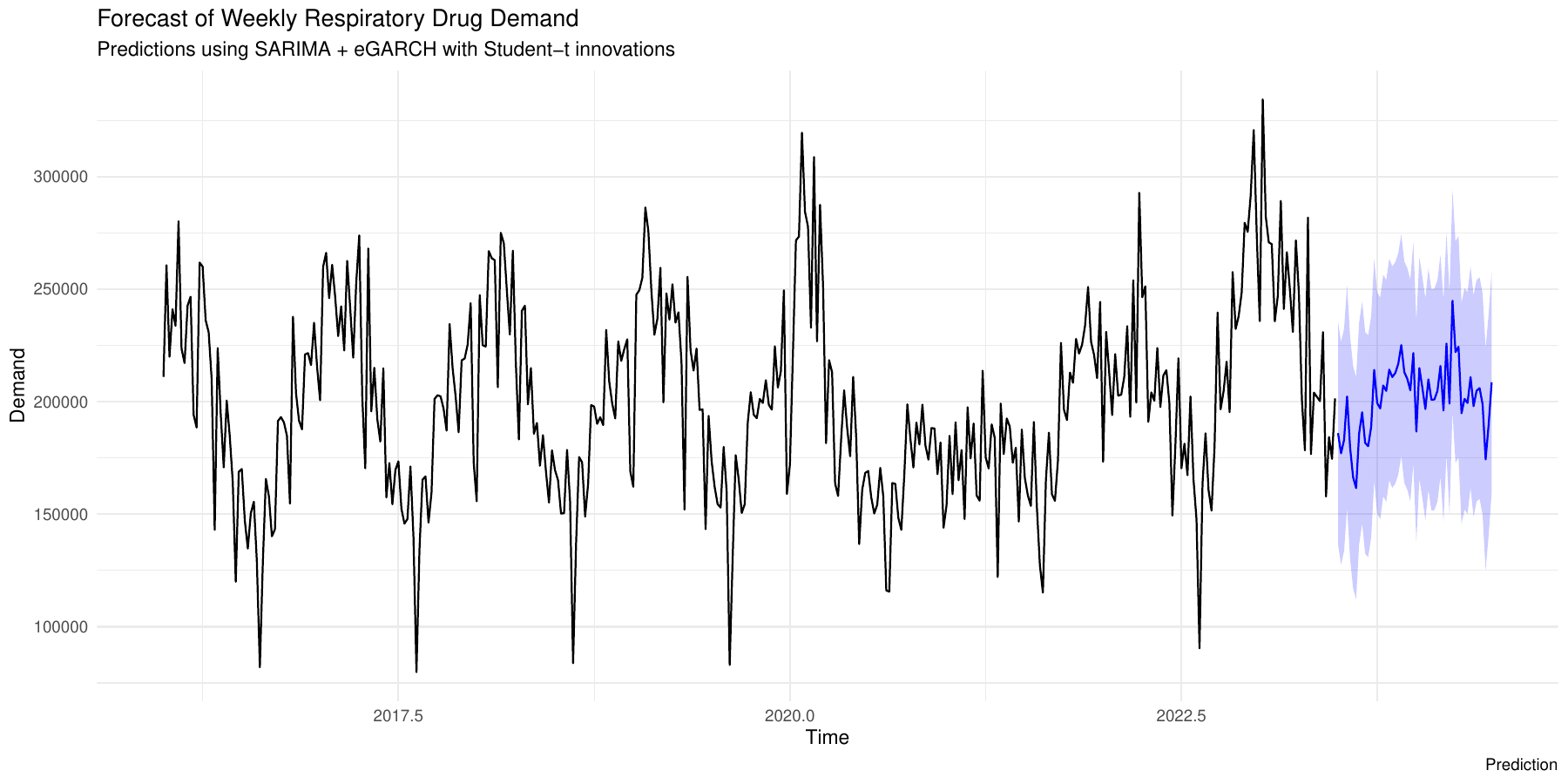}
    \caption{\textit{Forecast of weekly respiratory drug demand over a one-year horizon, obtained using a $\text{SARIMA}(4,0,0)(1,0,0)_{52}$ model estimated on the training set (excluding the final year). Volatility-adjusted confidence intervals are derived from an $\text{eGARCH}(1,1)$ model fitted to the SARIMA residuals, assuming Student-$t$ innovations. Predictive intervals incorporate the conditional standard deviations from the eGARCH model, accounting for both deterministic seasonality and time-varying volatility.}}
    \label{fig:sarima_garch_forecast}
\end{figure}

The predictive accuracy of the SARIMA–eGARCH model is noteworthy, although slightly lower than that of the univariate benchmark model Prophet. The MAPE of 14.9566\% indicates reasonably accurate forecasts for practical purposes. The RSR value of 0.8955 suggests that the prediction error remains below 90\% of the standard deviation of the observed values, reflecting good overall performance. 

\section{Holiday Effects in Prophet Forecasting}
\label{holiday_test}
Although Prophet allows for the inclusion of holiday effects through user-specified regressors, these variables were not incorporated in this study. A Welch two-sample t-test comparing average pharmaceutical demand during Christmas and New Year weeks with all other weeks yielded a non-significant result ($p$-value = 0.2185), with a 95\% confidence interval that spans both negative and positive values (CI: $[-8545.7,\ 34783.2]$). This indicates that any potential holiday-related effects are not statistically distinguishable from random variation.\\
Moreover, preliminary experiments showed that including holiday regressors did not improve predictive performance and, in some cases, slightly degraded it, likely due to overparameterization and the introduction of noise in the trend component. Since Prophet forecasts are used as input variables in all the other models presented, we opted for a simpler and more parsimonious specification to ensure consistency and avoid propagating unnecessary model complexity.

\section{Additional VARX Diagnostics and Justification}
\label{VAR_additional}
A Johansen cointegration test was performed to further assess the long-run relationships among the endogenous variables. The results confirm the presence of two cointegrating relationships ($r=2$), which supports the estimation of the VARX in levels, as the cointegration matrix has full rank~\citep{lutkepohl2005new}. 

\begin{table}[htbp]
\centering
\caption{Johansen cointegration test results using the trace statistic. The model excludes deterministic trends in the cointegration relation, though an intercept is estimated outside the cointegration space.}
\label{tab:johansen}
\begin{tabular}{lrrrr}
\toprule
\textbf{Null hypothesis} & \textbf{Test Statistic} & \textbf{10\%} & \textbf{5\%} & \textbf{1\%} \\
\midrule
$r \leq 1$ & 69.03  & 7.52  & 9.24  & 12.97 \\
$r = 0$    & 143.17 & 17.85 & 19.96 & 24.60 \\
\bottomrule
\end{tabular}
\end{table}

Results in Table~\ref{tab:johansen} supports estimating the VARX in levels, as it confirms the existence of long-run equilibrium relationships among the variables. Differencing in such a context would remove the cointegrated structure and discard valuable information on persistent dynamics.

Adopting a specification in levels is particularly suitable here, as the goal is to explore how environmental variables influence pharmaceutical demand over extended horizons. Capturing long-run effects is crucial when investigating phenomena such as climate change, whose impact on public health—particularly regarding respiratory medication use—unfolds gradually over time.

FEVD results in Table~\ref{tab:fevd} show that drug demand is predominantly explained by its own past values. Among environmental factors, temperature emerges as the most relevant predictor, explaining up to 6.05\% of the forecast error variance. In contrast, wind speed and precipitation contribute marginally, providing consistency with the insights from both the MBB-RF importance and Granger causality analyses.

\begin{table}[htbp]
\centering
\caption{Bootstrap-based FEVD estimates for respiratory drug demand. Values represent the proportion of forecast error variance explained by temperature and past drug demand at different horizons (in weeks). The table shows the mean estimates and the corresponding 95\% confidence intervals obtained from 1,000 bootstrap replications.}
\label{tab:fevd}
\begin{tabular}{cccccc}
\toprule
\textbf{Horizon (weeks)} & \textbf{Source} & \textbf{Mean} & \textbf{Lower Bound} & \textbf{Upper Bound} \\
\midrule
\multirow{2}{*}{4}  & Temperature       & 0.0574 & 0.0200 & 0.1016 \\
                    & Drug Demand       & 0.9426 & 0.8984 & 0.9800 \\
\midrule
\multirow{2}{*}{8}  & Temperature       & 0.0599 & 0.0222 & 0.1035 \\
                    & Drug Demand       & 0.9401 & 0.8965 & 0.9778 \\
\midrule
\multirow{2}{*}{12} & Temperature       & 0.0604 & 0.0223 & 0.1037 \\
                    & Drug Demand       & 0.9396 & 0.8963 & 0.9778 \\
\midrule
\multirow{2}{*}{16} & Temperature       & 0.0604 & 0.0223 & 0.1037 \\
                    & Drug Demand       & 0.9396 & 0.8963 & 0.9778 \\
\midrule
\multirow{2}{*}{20} & Temperature       & 0.0605 & 0.0223 & 0.1039 \\
                    & Drug Demand       & 0.9395 & 0.8961 & 0.9777 \\
\midrule
\multirow{2}{*}{24} & Temperature       & 0.0605 & 0.0223 & 0.1039 \\
                    & Drug Demand       & 0.9395 & 0.8961 & 0.9777 \\
\midrule
\multirow{2}{*}{28} & Temperature       & 0.0605 & 0.0223 & 0.1039 \\
                    & Drug Demand       & 0.9395 & 0.8961 & 0.9777 \\
\midrule
\multirow{2}{*}{32} & Temperature       & 0.0605 & 0.0223 & 0.1039 \\
                    & Drug Demand       & 0.9395 & 0.8961 & 0.9777 \\
\midrule
\multirow{2}{*}{36} & Temperature       & 0.0605 & 0.0223 & 0.1039 \\
                    & Drug Demand       & 0.9395 & 0.8961 & 0.9777 \\
\midrule
\multirow{2}{*}{40} & Temperature       & 0.0605 & 0.0223 & 0.1039 \\
                    & Drug Demand       & 0.9395 & 0.8961 & 0.9777 \\
\midrule
\multirow{2}{*}{44} & Temperature       & 0.0605 & 0.0223 & 0.1039 \\
                    & Drug Demand       & 0.9395 & 0.8961 & 0.9777 \\
\midrule
\multirow{2}{*}{48} & Temperature       & 0.0605 & 0.0223 & 0.1039 \\
                    & Drug Demand       & 0.9395 & 0.8961 & 0.9777 \\
\midrule
\multirow{2}{*}{52} & Temperature       & 0.0605 & 0.0223 & 0.1039 \\
                    & Drug Demand       & 0.9395 & 0.8961 & 0.9777 \\
\bottomrule
\end{tabular}
\end{table}

\section{Rolling window forecasting for robustness assessment}
A rolling window validation framework was implemented to evaluate the temporal robustness and generalization performance of the proposed forecasting models. This approach sequentially trains each model on a moving window of 260 weeks (five years) and forecasts the subsequent 52 weeks. Performance metrics were computed at each step and averaged over all iterations. This design allows us to assess how models behave under evolving data distributions and to detect potential overfitting, degradation, or instability over time.

A key objective of this analysis is to examine the impact of incorporating Prophet-fitted values as external regressors. For each model, two versions were tested: one including Prophet forecasts for drug demand and temperature, and one excluding them.

Results reveal that the standalone Prophet model performs poorly under rolling validation. Its MAPE rises to 20.27 and RSR reaches 1.34, indicating that the model struggles to adapt when faced with substantial structural changes in the data. This is likely due to the limited length of the training windows, which may not provide enough historical context for Prophet to capture shifts in trend or seasonality adequately. As a result, its forecasts fail to generalize effectively—highlighting Prophet’s limited capacity to accommodate abrupt regime shifts or evolving seasonal patterns, due to its firm reliance on smooth, predefined components that are not re-estimated adaptively.

More critically, when Prophet-fitted values are included as external inputs, they consistently degrade model performance in the rolling window framework. For the VARX model, MAPE increases from 12.05 (without Prophet) to 22.30 (with Prophet), and RSR jumps from 0.8350 to 1.44. Similarly, for LSTM, MAPE increases from 14.03 to 20.01, and RSR from 0.9306 to 1.21. These substantial degradations can largely be attributed to the poor standalone performance of Prophet on individual slices. When Prophet's forecasts are inaccurate or unstable, using them as exogenous inputs introduces noise into the model, thereby reducing its ability to adapt and generalize. This effect is particularly pronounced in settings with structural variability, where Prophet struggles to capture abrupt or irregular changes.

The MBB-RF model shows greater resilience, with MAPE worsening only from 10.84 to 12.38 and RSR from 0.7623 to 0.8786. This robustness stems from the model’s capacity to down-weight or disregard uninformative predictors during tree construction, thereby mitigating the negative impact of inaccurate Prophet inputs.

While all models exhibit a general decline in performance under rolling validation—due to reduced training set size and increased exposure to temporal heterogeneity—the comparison between Prophet-augmented and non-augmented versions, under identical data conditions, clearly shows that excluding Prophet yields better forecasting accuracy.

These results suggest that Prophet-fitted values should not be included as exogenous inputs by default in rolling window settings. Instead, we recommend incorporating them only when Prophet demonstrates superior standalone performance in a standard forecasting scenario, such as the forward holdout evaluation conducted in our study. This ensures that Prophet contributes meaningful information, rather than introducing spurious structure. In our case, although Prophet performed well in the holdout setting, its forecasts failed to generalize under more demanding rolling conditions—highlighting the importance of context-aware validation when designing hybrid forecasting systems.

Notably, across all configurations, the MBB-RF model emerges as the most robust performer in the rolling window evaluation. Compared to LSTM, its nonparametric nature allows it to achieve strong performance even with limited training data, without the need for extensive sequence modeling. In contrast to VARX, MBB-RF is better equipped to handle non-linear relationships and high-order interactions among features, and its use of block bootstrap enhances its resilience to temporal dependence. These characteristics make MBB-RF particularly well suited for complex, real-world time series settings with evolving dynamics and limited stationarity.

\begin{figure}[h!]
    \centering
    \includegraphics[scale = 0.5]{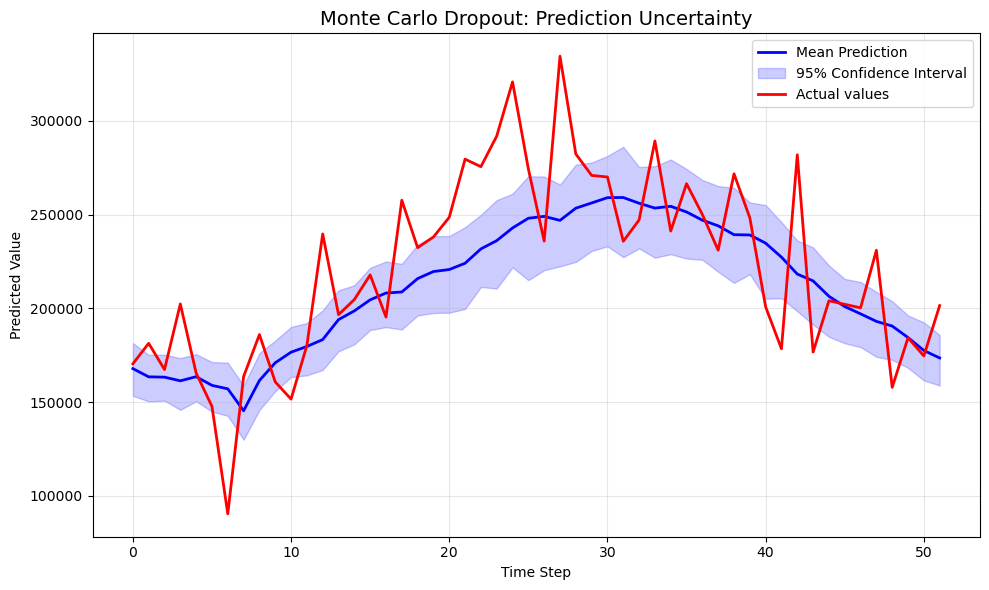}  
    \caption{\textit{Monte Carlo Dropout-based uncertainty estimation for LSTM model. The plot shows the predicted mean (blue), 95\% confidence intervals (shaded area), and observed values (red).}}
    \label{fig:mc_dropout}
\end{figure}

\section{LSTM Stability Analysis}
\label{lstm}
A 20\% dropout rate was reactivated during inference using Monte Carlo Dropout to evaluate the stability and reliability of the LSTM forecasts. Unlike conventional inference procedures where dropout is turned off, this method estimates model uncertainty by running the same input through the network 1,000 times with stochastic dropout enabled. The resulting distribution of predictions allows for the computation of confidence intervals around the mean forecast. Narrow confidence bands reflect stable and robust model behavior. In contrast, wider intervals may indicate greater sensitivity to input variations or potential instability in the predictive process.

The smooth and mostly narrow confidence bands in Figure~\ref{fig:mc_dropout} suggest that the model predictions are generally stable. The localized widening of the bands aligns with periods of increased variability in the observed data.

\section*{References}
Dagum, E. B. (2001). Analisi delle serie storiche: modellistica, previsione e scomposizione. Springer Science \& Business Media.

Francq, C. and Zakoian, J.-M. (2019). GARCH models: structure, statistical inference and financial applications. John Wiley \& Sons. 

Lütkepohl, H. (2005). New introduction to multiple time series analysis. Springers Science \& Business Media.

Nelson, D. B. (1991). Conditional heteroskedasticity in asset returns: A new approach. Econometrica: Journal of the econometric society, pages 347–370.

\end{document}